\documentclass[%
 aip,
 amsmath,amssymb,
 reprint,%
]{revtex4-1}

\usepackage{graphicx}
\usepackage{dcolumn}
\usepackage{bm}
\usepackage{comment}
\usepackage[utf8]{inputenc}
\usepackage[T1]{fontenc}
\usepackage{mathptmx}
\usepackage{etoolbox}
\usepackage[caption=false]{subfig}
\usepackage{floatrow}

	\global\long\def\grad{\nabla}%
	\global\long\def\elaspush{\mathbf{C}}%
	\global\long\def\bm{\bn m}%
	\global\long\def\disp{\mathbf{u}}%
	\global\long\def\pt{\mathscr{PT}}%
	\global\long\def\inner#1#2{\left\langle #1,#2\right\rangle }%
	\global\long\def\T{\mathbf{T}}%
	\global\long\def\pt{\mathcal{PT}}%
	\global\long\def\op{\mathcal{A}}%
	\global\long\def\mode{\phi}%
	\global\long\def\op{\mathcal{H}}%

\global\long\def\inner#1#2{\langle#1,#2\rangle}%

\def\bm#1{\mathbf{#1}}

\makeatletter
\def\@email#1#2{%
 \endgroup
 \patchcmd{\titleblock@produce}
  {\frontmatter@RRAPformat}
  {\frontmatter@RRAPformat{\produce@RRAP{*#1\href{mailto:#2}{#2}}}\frontmatter@RRAPformat}
  {}{}
}%
\makeatother
\begin{document}

\preprint{AIP/123-QED}

\title[Perspective on Non-Hermitian elastodynamics]{Perspective on Non-Hermitian Elastodynamics}

\author{Johan Christensen}%
\affiliation{IMDEA Materials Institute, Calle Eric Kandel, 2, 28906 Getafe, Madrid, Spain
}%

\author{Michael R.~Haberman}%
\affiliation{Walker Department of Mechanical Engineering, The University of Texas at Austin, Austin, TX 78712-1591, USA}

\author{Ankit Srivastava}%
\affiliation{Department of Mechanical, Materials, and Aerospace Engineering
Illinois Institute of Technology, Chicago, IL, 60616 USA}

\author{Guoliang Huang}%
\affiliation{Department of Mechanical and Aerospace Engineering, University of Missouri, Columbia, MO 65211, USA}

\author{Gal Shmuel}
\affiliation{
Faculty of Mechanical Engineering, Technion–Israel Institute of Technology, Haifa 32000, Israel%
}%
\email{johan.christensen@imdea.org, meshmuel@technion.ac.il}

\date{\today}

\begin{abstract}
The manipulation of mechanical waves is a long-standing challenge for scientists and engineers, as numerous devices require their control. 
The current forefront of research in the control of classical waves has emerged from a seemingly unrelated field, namely, non-Hermitian quantum mechanics. By drawing analogies between this theory and those of classical systems, researchers have discovered phenomena that defy conventional intuition and have exploited them to control light, sound, and elastic waves.
Here, we provide a brief perspective on recent developments, challenges and intricacies that distinguish non-Hermitian elastodynamics from optics and acoustics. We close this perspective  with an outlook on potential directions such as topological phases in non-Hermitian elastodynamics and broken Hermitian symmetry in materials with  electromomentum couplings. 
\end{abstract}

\maketitle

\section{\label{sec:intro}Introduction}

The past two decades have shown that the properties of artificial materials can be tailored to exhibit extraordinary dynamic behavior and properties by cleverly engineering their composition and structure. The development of such  metamaterials is a prominent thrust in engineering today \cite{cummer2016controlling, haberman2016acoustic, kochmann2017exploiting, Kadi2019nrp,Bigoni2013PRB,Christensen2015MRCComunications,gonella15,kochmann14pre,parnell2012employing}. The principle focus of metamaterial research is tailored wave control based on the design of subwavelength structure. Elastic waves are of particular interest given the numerous mechanical applications that require their control, such as vibration isolation, impact mitigation, ultrasonography, energy harvesting, and stealth, to name just a few.

Currently, the forefront of research in the control of classical wave motion emerged from a seemingly unrelated field, namely, quantum mechanics, with the development of its non-Hermitian formalism\cite{bender1998real,MOSTAFAZADEH2010ijgmp,moiseyev2011book}. 
This formalism describes open quantum systems that exchange energy with their environment, resulting with non-orthogonal or even colinear natural modes and degenerate eigenvalues in contrast with Hermitian systems whose natural modes are orthogonal.  
By drawing analogies between this formalism and those of classical systems\cite{Feng2017np,El-Ganainy2018ys,Ozdemir2019cr,Miri2019science,Pile:2017jk,Midya2018natcom,El-Ganainy2019cp}, researchers have discovered phenomena that defy conventional intuition and have used them to control light \cite{el2007theory,klaiman2008visualization,guo2009observation,ruter2010observation,Longhi2017,Goldzak2018PRL,Zhao2018nsr}, sound \cite{zhu2014p,Fleury2015,Cummer2016,shi2016accessing, xu2020prl,Thevamaran2019},
and elastic waves \cite{Christensen2016prl,hou2018jap,Merkel2018,Hou2018PRApplied,Psiachos2018jap,Merkel2018prb}. Of all the  branches of classical physics, these concepts were least studied in elastodynamics, in spite of (or potentially due to) the fact that elastodynamics exhibits a distinct tensorial richness. In this perspective, we discuss the intricacies and challenges that distinguish non-Hermitian elastodynamics from optics and acoustics, review some of the recent developments in this field, and present an outlook to potential directions from the current state-of-the-art.  

\section{\label{sec:background}Background}

\begin{figure}[t!]
	\centering\sidesubfloat[]{\includegraphics[width=.6\textwidth]{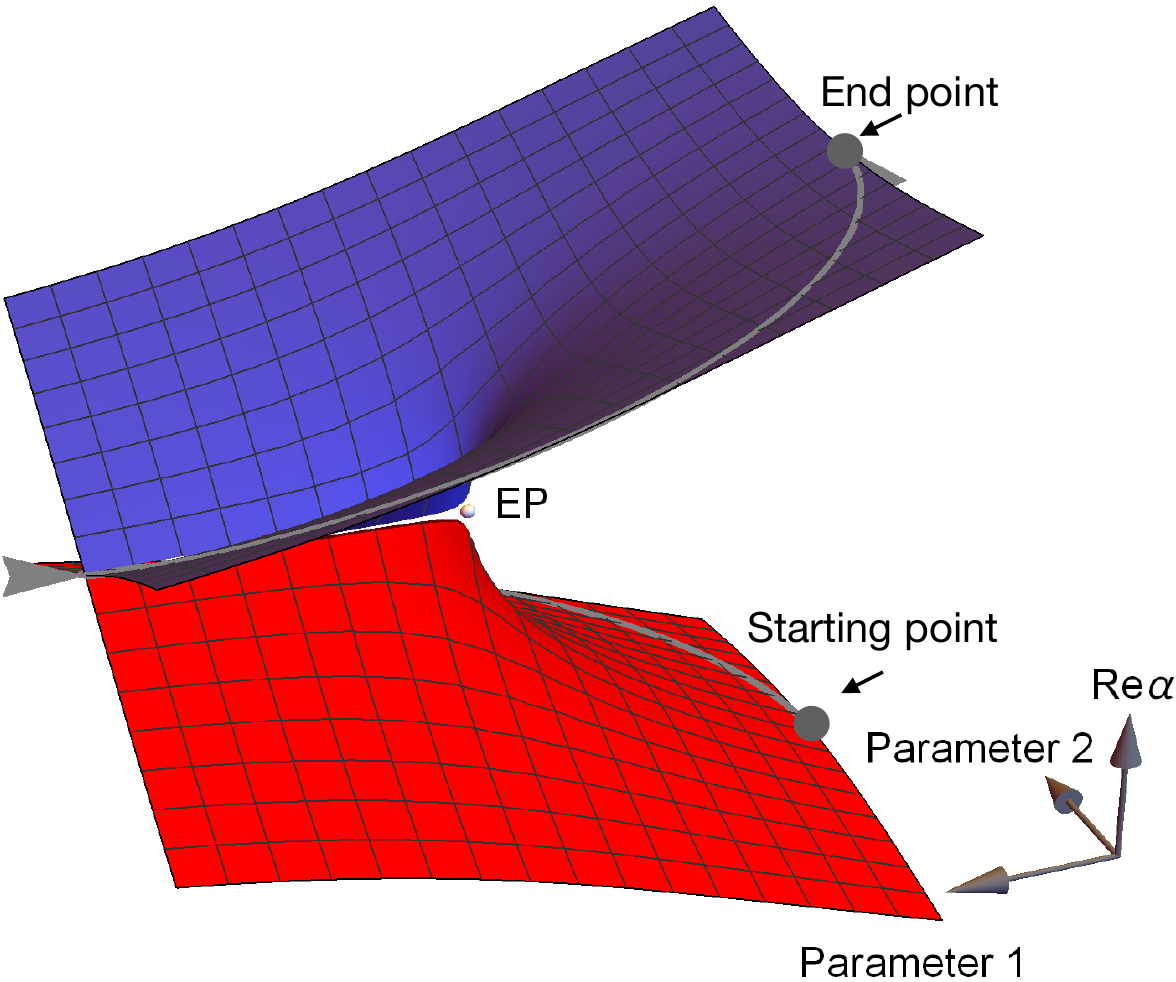}}  \\
 \sidesubfloat[]{\includegraphics[width=.6\textwidth]{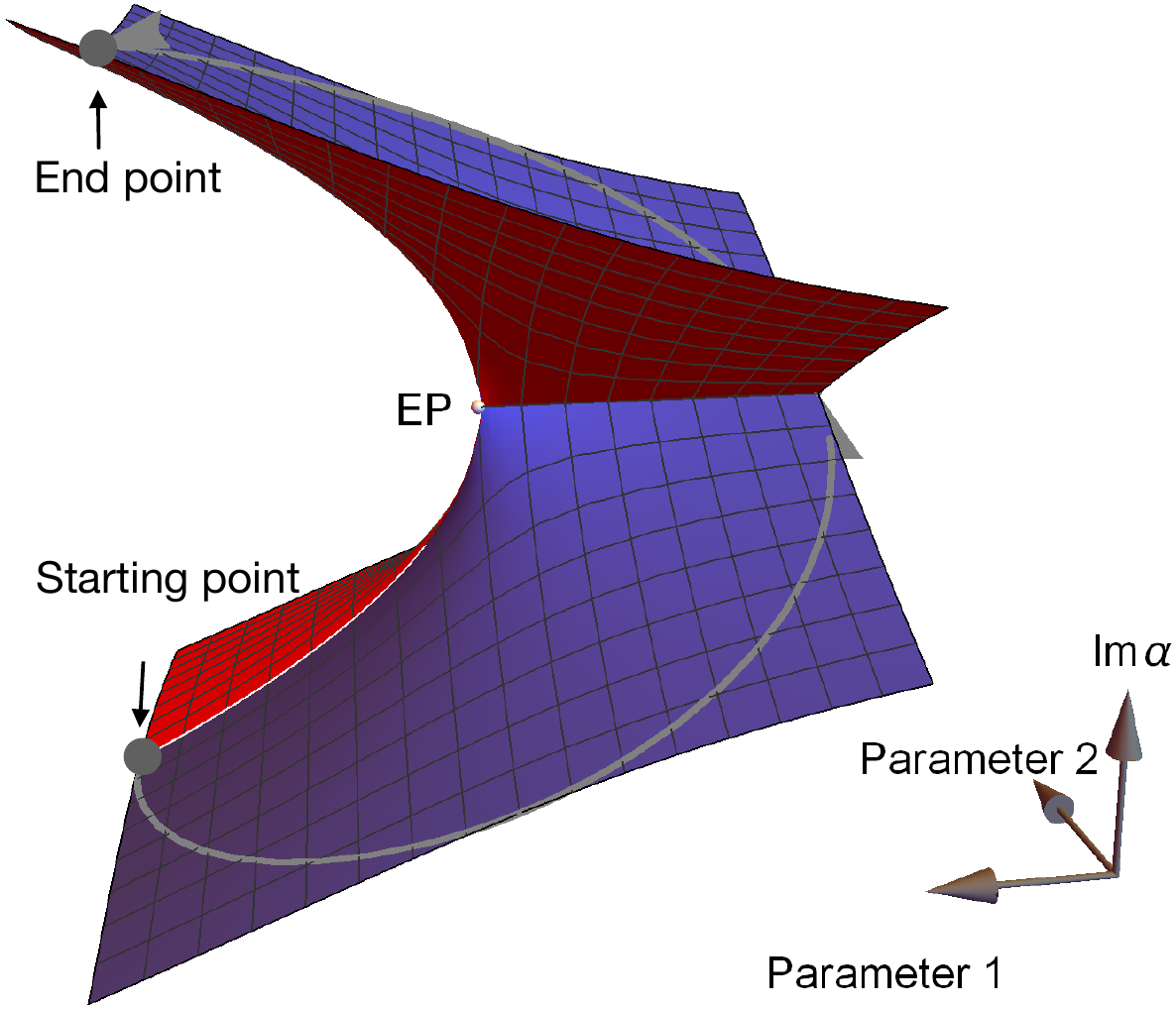}}  \caption{The (a) real and (b) imaginary parts of the eigenvalues as functions of two suitable parameters of a non-Hermitian operator. For critical values of the two parameters, the Riemann sheets degenerate to a so-called exceptional point. The gray curves depict a loop around the EP which results in a different state, owing to the multi-valued nature of the Riemann sheet. Conventional research on non-Hermitian systems considers the natural frequencies as the eigenvalues and requires material gain (or loss) to be one of the parameters. In a recent paradigm shift\cite{Lustig2019,mokhtari2020scattering,FISHMAN2024JMPS}, the EPs are designed in the wavevector space, using the unique  coupling of elastic wave polarizations. Thus, the EPs are formed without material gain or loss, and the parameters that control the wavevector are the tuning parameters.}
	
	{\small{}{}\label{fig:epx}}{\small\par}
\end{figure}

Eigenvalue problems are ubiquitous in physics: they describe the natural modes of the relevant system using a suitable operator. Formally, they may be written as
$
\op\mode=\alpha\mode$
where $\op$ is an operator acting on the natural mode $\mode$, 
generating its multiplication by the (possibly complex) eigenvalue
$\alpha$. In quantum mechanics, the operator is called the Hamiltonian, which operates on the wave function in the Schrödinger equation.
One of the fundamental postulates in quantum mechanics is that the Hamiltonian exhibits a mathematical symmetry termed `Hermiticity' which ensures real eigenvalues and is associated with energy conservation\cite{Chew2008,Srivastava2015prsa,pernassalomn2020prapplied}. On a basic level, Hermiticity implies that for any two functions $\mode$ and $\varphi$ in a function space endowed with an inner product $\inner{\cdot}{\cdot}$, the Hermitian operator $\op$ satisfies
\begin{equation}
\inner{\op\mode}{\varphi}=\inner{\mode}{\op\varphi}.
\end{equation}
Strikingly,  \citet{bender1998real} discovered
that Hamiltonians  with so-called $\pt$ symmetry
may also support real spectra. 
Unlike the eigenmodes of Hermitian operators, the eigenmodes of  $\pt$-symmetric operators are no longer orthogonal one to another. In the extreme case, these modes  may even coalesce together with their eigenvalues at the so-called `exceptional points' (EPs) \cite{Longhi2015SR,Longhi2017jpa,Longhi2017}. These non-Hermitian degeneracies arise when suitable parameters of non-Hermitian systems are appropriately tuned. Near EPs, the functional dependency of the eigenvalues on these parameters defines self-intersecting Riemann sheets, as shown in Fig.\ \ref{fig:epx}.

\floatsetup[figure]{style=plain,subcapbesideposition=top}

EPs and the unique topology around them are the source of fascinating counterintuitive phenomena such as chiral modes\cite{Peng2016pnas}, supersensitivity\cite{hodaei2017enhanced} and unidirectional zero reflection \cite{Shen2018prmat}.
While the bulk of the research (and the above discussion) considers $\pt$-symmetric Hamiltonians, they belong to a larger class of non-Hermitian Hamiltonians\cite{Mostafazadeh2002jmp,MOSTAFAZADEH2010ijgmp,Suchkov2016njp,Lustig2019} that can exhibit real eigenvalues and EPs at critical values of a suitable parameter set. Understanding that these phenomena rely upon the nature of eigenvalue problems associated with the wave operator has led to their dissemination in other wave physics. We summarize next the fundamental operators in the different classical systems in order to highlight the uniqueness of non-Hermiticity in elastodynamics.

In photonics, Maxwell's equations can be expressed in the form of an eigenproblem \cite{griffiths1999introduction,Joannopoulos2008book} for the magnetic eigenmodes $\mathbf{H}$ and their frequencies $\omega^{2}$ (the electric field can be determined subsequently). The corresponding wave operator is a function of the second-order dielectric tensor $\boldsymbol{\epsilon}$ (since the permeability of most materials is very close to the vacuum permeability\cite{Joannopoulos2008book}). Magnetic waves are subjected to the constraint $\grad\cdot\mathbf{H}=0,$ hence do not support longitudinal modes and are classified as transverse waves.

In acoustics, i.e., in the study of waves in gases and liquids without resistance to shear deformation\cite{kinsler2000fundamentals}, the eigenvalue problem can be expressed as a scalar equation for the pressure field, whose operator depends on the mass density¸ $\rho$, and bulk modulus, $K$, of the fluid in which waves propagate\cite{kinsler2000fundamentals}. The velocity field $\mathbf{v}$ of acoustic waves is subjected to the constraint $\grad\times\mathbf{v}=\mathbf{0}$, and thus do not support transverse modes. Acoustic waves are longitudinal pressure waves, which are also referred to as dilatational or volumetric waves\cite{graff1975wave}, since they are accompanied with volume change, in contrast to transverse waves, i.e., shear waves which are isochoric.

In solid mechanics, the elastodynamics equations
can be formulated as an eigenvalue problem for the time- and space- dependent displacement vector field $\mathbf{u}(\mathbf{x},t)$ of material points (see Appendix \ref{app:non-herm}).
The wave operator is a function of the mass density and the elasticity tensor of the solid, $\elaspush$. Its simplest form is the Christoffel equation\cite{auld1973acoustic,nayfeh1995wave} for bulk plane waves in the $\mathbf{n}$ direction: $k^2\rho^{-1}\boldsymbol{\Gamma}(\disp)=\omega^2 \disp$ where $k$ is the wavenumber and the components of the Christoffel operator are $\Gamma_{il}=C_{ijkl}n_j n_l$. The first source of richness originates from the dimension of $\elaspush$, which is a fourth order tensor and thus constitutes a larger design space in comparison with photonics and acoustics. 
Second, unlike the magnetic field (subjected to $\grad\cdot\mathbf{H}=0$) and the velocity field in acoustics (subjected to $\grad\times\mathbf{v}=\mathbf{0}$) the displacement field $\mathbf{u}\left(\mathbf{x},t\right)$ is not subjected to any differential constraint. Thus, even the simplest solids (conservative, isotropic and homogeneous) 
support both transverse and longitudinal wave polarizations. These inherent modes  can be coupled one with the other, a coupling that gives rise to unique physics \cite{Long2018pnas}. 

Most works on non-Hermitian physics focus on problems of the type mentioned above, in which the frequency (squared) is the eigenvalue and corresponds to the propagation of free waves owing to some impulsive excitation, i.e., in the absence of a constant excitation. Assuming real wavenumbers (no radiation losses), the breaking of Hermitian symmetry  (i.e., accessing complex frequencies and nonorthogonal eigenmodes) in these problems requires material loss or gain. 
While optical gain is well-established using stimulated emission, generating and controlling elastic gain remains a challenge. We list the recent advances for tackling this challenge in the first part of Sec.~\ref{sec:state of the art}, together with their applications.

Less explored from the perspective of non-Hermitian physics, a dual class of problems involves a formulation where the wavenumber is the eigenvalue and the frequency is a prescribed real quantity (see Appendix \ref{app:non-herm}). This formulation pertains to conservative media with open boundaries and sustained driving sources. Recent progress has utilized the unique coexistence and coupling of wave polarizations in elastodynamics to engineer EPs in such conservative solids, thus circumventing the challenges associated with the design of material gain and loss\cite{Lustig2019,mokhtari2020scattering,FISHMAN2024JMPS}. The second part of Sec.~\ref{sec:state of the art} expands on the works that introduced this approach and how they capitalized on emergent non-Hermitian features to manipulate elastic waves.    



\section{\label{sec:state of the art}Recent developments}

As discussed in Sec.\ \ref{sec:background}, the common approach for breaking Hermitian symmetry requires the careful design of material loss and gain. Piezoelectric materials are therefore natural candidates to achieve non-Hermitian behavior in elastodynamic systems due to their ability to convert electric energy to mechanical energy and vice versa. 
In a series of works \cite{Christensen2014prb, Christensen2015acta, Christensen2016epl,gao2020prl}, Christensen and collaborators developed a procedure that exploits the acousto-electric effect in piezoelectric semiconductors for material gain, thereby synthesizing $\mathcal{PT}$-symmetric elastic media \cite{Christensen2016prl} and breaking elastodynamic reciprocity \cite{Merkel2018}. When an acoustic field impinges on a piezoelectric semiconductor slab a coherently oscillating electric charge is created. Superposing a sufficiently high DC electric field corresponding to a supersonic carrier drift speed leads to sound amplification by virtue of phonon emission, an effect known as acoustic Cherenkov radiation. In particular, this acousto-electric effect has been employed for two demonstrations. First, stacking multiple piezoelectric semiconductors was shown to generate a system with balanced loss and gain with associated non-Hermitian properties, if the layers are adequately electrically loaded \cite{Christensen2016prl}. Next, a theoretical non-Hermitian Su-Schrieffer-Heeger demonstration was made using the same approach. In that case, the non-Hermitian skin effect and the ensuing failure of both the Bloch band topology were demonstrated \cite{gao2020prl}. We note, however that experimental demonstrations of the latter theoretical prediction has not yet been provided.

A more conventional approach to use piezoelectric materials is to shunt them through electrical circuits of resistors, capacitors, and inductors \cite{GARDONIO2017}, which collectively generate tunable elastic gain loss\cite{hou2018jap,GARDONIO2017}. Assouar’s group proposed a tunable $\pt$-symmetric elastic beam based on such shunted piezoelectric elements\cite{hou2018jap}. They later employed this concept for negative refraction of flexural waves\cite{Hou2018PRApplied}. Similarly, Huang's group employed shunted piezoelectric patches attached to a $\pt$-symmetric elastic beam that displayed asymmetric flexural wave scattering\cite{wu2019asymmetric}. 

Ruzzene and colleagues have also used shunted piezoelectric arrays to generate spatiotemporal modulation of the elastic modulus, thereby breaking reciprocal elastic wave transport for flexural waves\cite{marconi2020prapplied,Xia2021prl}.  This approach was later employed by Erturk and colleagues.\cite{Thomes2024apl} to demonstrate the formation of a resonant EPs. Such materials that violate time invariance can exhibit a wide array of nonreciprocal wave phenomena such as one-way wave amplification and attenuation\cite{nassar2017modulated,torrent2018loss,nassar2020nonreciprocity}. From a modeling perspective, the constitutive parameters and phase velocity become time-dependent. 
These works specifically considered modulations in the form of progressive periodic waves ('pump waves'), 
which  create a spatiotemporal bias and nonreciprocal wave motion as a function of the modulation. 
Small-amplitude medium-speed modulations lead to a Bragg scattering regime; faster modulations can be described by an equivalent medium with effective properties; and slower but higher-amplitude modulations lead to an adiabatic regime\cite{nassar2020nonreciprocity}. This nonreciprocal wave propagation is fundamentally different from one-way Bragg reflection, but can still be leveraged for the purposes of one-way mirroring, especially at low frequencies where wide bandgaps are seldom available. In this scenario, Hooke’s law no longer applies 
and must be replaced by constitutive relations of the Willis form\cite{willis1981avariational,willis2011effective,WILLIS2012MRC}, which state that the stress and linear momentum depend both the strain \emph{and} velocity fields. The Willis parameters in the case of spatiotemporal modulations capture the nonreciprocal nature of the modulated microstructures\cite{nassar2017modulated}. Indeed, Willis materials offer another platform for breaking the Hermitian nature\footnote{Willis uses the term self-adjoint\cite{ willis2011effective,WILLIS2012MRC} rather than Hermitian.}, as we discuss later.

Collectively, the works above have tackled the important challenge of generating gain in elastic materials in order to break temporal hermiticity. 
However, gain is not a necessity for realizing EPs and some of their associated features are accessible using loss alone, an inherent feature of viscoelastic materials. Indeed, \citet{shmuel2020prapplied} showed that the introduction of a viscoelastic part in an elastic slab may generate EP of two modes (EP2) in the spectrum of the assembly. They applied non-Hermitian perturbation theory previously developed in quantum mechanics to determine the conditions under which this occurs and employed the resultant topology around the EP for mass sensing with enhanced sensitivity. Similarly, \citet{rocha2020PRApplied} used differential loss instead of gain and loss to form EP2. While the modes they analyzed were structural modes (torsional modes of pillars) rather than elastodynamics modes, their work is one of the few experimental demonstrations of the sensitive frequency splitting near EP of a mechanical system.

\begin{figure}[t!]
{\includegraphics[width=.9\textwidth]{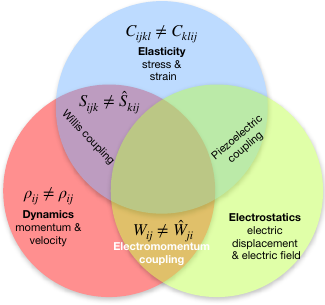}}
 \caption{Possible non-Hermitian constitutive operators in elastodynamics. Elastic materials may exhibit odd elastic tensor ($\mathbf{C}=-\mathbf{C}^\text{T}$), Willis materials may exhibit odd Willis- ($\hat{\mathbf{S}}=-\mathbf{S}^\text{T}$)  and density ($\boldsymbol{\rho}=-\boldsymbol{\rho}^\text{T}$) tensors, and electromomentum materials may exhibit odd electromomentum tensors ($\hat{\mathbf{W}}=-\mathbf{W}^\text{T}$).   }
	
	{\small{}{}\label{fig:odd}}{\small\par}
\end{figure}

The nonconservative nature of the media presented so far manifests itself through time-dependent constitutive equations and hence complex elastic moduli. These, in turn, yield complex frequencies for the Christoffel equation. Another type of nonconservative media are Cauchy-elastic media\cite{ogden97book}: the work done by the stress in such media generally depends on the deformation path, and therefore cannot be derived from a strain energy function. As a result, the elasticity tensor no longer exhibits the major symmetry that is needed to ensure real frequencies in the Christoffel equation.  Under the name 'odd elasticity', Vitelli and colleagues\cite{scheibner2020odd,Scheibner2020prl} set forth a model of active material with nonconservative microscopic interactions that generate odd elastic moduli, and studied the non-Hermitian elastodynamics of such materials with odd elasticity (top circle in Fig.\ \ref{fig:odd}). 
They demonstrated that in isotropic solids with odd elasticity, the wave polarizations (eigenmodes) are no longer orthogonal, and may even become colinear for a threshold value of the elastic moduli. This EP marks the transition to odd-elastic waves with circular polarization\cite{scheibner2020odd}.

Active material systems displaying odd elasticity have been experimentally demonstrated by introducing piezoelectric elements and motors controlled by electrical circuits, or spinning networks into host media \cite{brandenbourger2019non, chen2021realization}. \citet{chen2021realization} 1D non-Hermitian metamaterial in which each unit cell consists of three piezoelectric patches mounted on a steel beam. The metabeam was assumed to have two modes of deformation including bending and shearing. These modes of deformation in turn induce a shear stress $\sigma$ and a bending moment, $M$. The crucial difference between this non-Hermitian beam and a traditional beam is the presence of internal energy sources that violate energy conservation. By designing the feedback to create non-reciprocal coupling between the elongation $s$ and shear $b$, the constitutive relation of the metabeam takes the form of odd elasticity
\begin{equation}\label{equ1}
\left[\begin{array}{c}
\sigma\\
M
\end{array}\right]=\left[\begin{array}{ll}
\mu & P \\
0 & B
\end{array}\right]\left[\begin{array}{l}
s \\
b
\end{array}\right],
\end{equation}
i.e., the electronic feedback between the piezoelectric patches induces a new modulus $P$, in addition to the shear and bending moduli $\mu$ and $B$, respectively. Because the energy differential is $\ \delta \psi = \sigma \delta s + M \delta b$, the asymmetric part of the matrix corresponds to a violation of Maxwell–Betti reciprocity. The parity-violating, nonconservative modulus $P$ also induces unidirectional wave amplification. For finite structures, these non-Hermitian systems with odd elasticity exhibit non-Hermitian 1D and 2D skin effect and non-Hermitian Rayleigh wave propagation \cite{chen2021realization, gao2022non}. The skin effect in the non-Hermitian system are defined as the case where bulk modes behave as the skin modes collapsing at the open boundaries. 
Recently, the design of active microstructure for a 2D non-Hermitian odd plate with odd density through feed-forward interactions was suggested to explore a series of unconventional wave phenomena when conventional Timoshenko plate mechanics meets with non-Hermiticity\cite{wang2023non,wu2023active}. Those works numerically and experimentally demonstrated nonreciprocal wave amplification and attenuation phenomena along with direction-dependent dispersion control of flexural waves in 2D odd plates.

The approaches mentioned so far to design non-Hermitian elastodynamics can be generalized using Willis materials\cite{willis1981avariational,willis1997dynamics,willis2011effective}, which exhibit a different conversion of strain energy and kinetic energy relative to conventional materials\cite{Sieck2017prb}.
This distinct mechanism is reflected by the Willis tensors $\mathbf{S}$ and $\mathbf{\hat{S}}$ and tensorial mass density $\boldsymbol{\rho}$ appearing in nonlocal constitutive equations, analogous to the bianisotropic equations in electromagnetism\cite{milton06cloaking,Sieck2017prb}. In the spatially local limit, the constitutive  (Milton-Briane-Willis\cite{milton06cloaking,Milton2020II}) equations take the form $\sigma_{ij}=C_{ijkl}u_{k,l}+S_{ijk}\dot{u}_k$ and $p_i=\rho_{ij} \dot{u}_j+\hat{S}_{ijk}u_{j,k}$. Extending the concept of odd elasticity to odd mass density and odd Willis tensors thus breaks Hermitian symmetry in the Christoffel equation, giving rise to complex frequencies. This translates to designing a Willis material such that one of the symmetries $\ \rho_{ij}=\rho_{ji},\  \hat{S}_{ijk}=S_{kij}$ is broken (left circle in Fig.\ \ref{fig:odd}).
Huang and collaborators\cite{wu2023active} designed and experimentally realized such active microstructure that gives rise to odd mass density.  They further demonstrated the formation of EPs of transition between stable and unstable waves, directional wave ampliﬁcation and non-Hermitian skin effect in the medium. In another collaboration\cite{Chen2020nc}, they experimentally realized odd Willis couplings in active flexural media using piezoelectric sensor–actuator pairs controlled with digital circuits,  giving rise to nonreciprocal wave propagation. Li and collaborators\cite{Liu2019prx} introduced loss to a Willis beam, and showed the formation of EP in the scattering matrix for flexural waves, at  which there is unidirectional zero reflection occurs.

Building upon these works probing gain and loss in elastic materials, \citet{Psiachos2018jap} showed that coupled transverse-longitudinal elastic waves scatter asymmetrically from alternating layers with gain and loss. Importantly, they later realized that the coupling between transverse and longitudinal  waves is sufficient to generate asymmetric scattering even without gain and loss\cite{Psiachos2019prb}.

\begin{figure}[t!]
	\centering\sidesubfloat[]{\includegraphics[width=.7\textwidth]{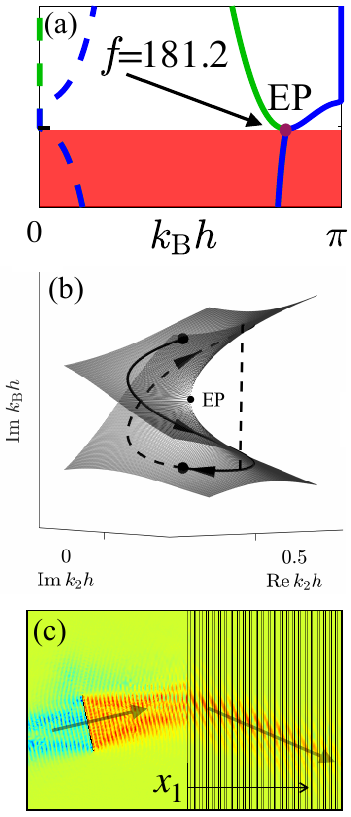}}
 \caption{\label{fig:eptwo} (a)  The coalescence of two Bloch modes of the exemplary conservative laminate considered by 
 \citet{Lustig2019}.  The EPs are shown in a cut in the  frequency-Bloch wavenumber $(f,k_\mathrm{B})$ diagram for normalized vertical wavenumber $k_{2}h=0.5$ about 181.2\,kHz. Re$\,k_\mathrm{B}$ and Im$\,k_\mathrm{B}$ are shown in solid and dashed curves, respectively. The real and imaginary parts of a certain branch are plotted using the same color. (b) One of the Riemann sheets associated with this EP in the wavevector space. The solid and dashed curves illustrate the asymmetric nature of a loop around the EP. One direction (solid curve) is adiabatic and the other (dashed curve) is non-adiabatic, resulting in a different state at the end of the loop. (c) Full-wave finite element simulations of the axial energy flux in the pertinent scattering problem near the EP. The energy flux is preserved in the laminate and exhibits negative refraction. See also supplementary video 1.}
	
	{\small{}{}\label{fig:laminateep}}{\small\par}
\end{figure}

All the developments that are listed above rely upon material gain, loss or their combination in order to access non-Hermitian features in elastodynamics. Recently, Shmuel's group\cite{Lustig2019} introduced a unique paradigm to eliminate the need for material gain or loss by breaking \emph{spatial} Hermitian symmetry using the tensorial nature that is unique to elastodynamics. From a mathematical perspective, they designed the non-Hermitian part of the operator for the wavenumbers instead of the operator for the natural frequencies (see Appendix \ref{app:non-herm}). From a mechanical perspective, the analysis is of a conservative solid with open boundaries and sustained driving source. More specifically, they revisited the canonical scattering problem of  monochromatic plane waves impinging on a semi-infinite periodic laminate made of two conservative isotropic materials\cite{joseph2015WM}. By designing the unit cell, they tuned the energy transfer between the two elastic polarizations at the material interfaces such that two of the forward Bloch modes coalesce\cite{Lustig2019} (Fig.\ \ref{fig:eptwo}a). \floatsetup[figure]{style=plain}
\captionsetup[subfloat]{position=top,labelformat=empty}

The EPs that are associated with this non-Hermitian eigenmode degeneracy are formed in the complex wavevector space, in contrast with the gain and complex frequency parameter space as in the works discussed above. These non-resonant EPs are surrounded by a self-intersecting Riemann surface  (Fig.\ \ref{fig:eptwo}b), whose curvature is the related to the energy flow and therefore the unique topology of these surfaces can be leveraged for anomalous energy transport. For example, \citet{Lustig2019} showed that these spatial EPs give rise to negative refraction (Fig.\ \ref{fig:eptwo}c), as they are the crossing of branches of opposite slopes. The key point is that, contrary to previous works discussed above, the parameters of the Riemann surface about the non-resonant EPs do not correspond to material gain or loss, but, e.g., the angle of the incident wave as depicted in the left half-space in Fig.\ \ref{fig:eptwo}c).   Subsequently, Srivastava's group\cite{mokhtari2020scattering} developed designated tools to further analyze the non-Hermitian operator, its EPs and scattering spectrum, suggesting its connection to resonance trapping. 

\begin{figure}[t!]
	\centering\sidesubfloat[]{\includegraphics[width=.8\textwidth]{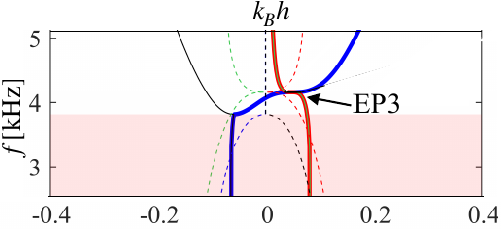}}  
 \bigskip
 \\

 \sidesubfloat[]
  {\includegraphics[width=.8\textwidth]{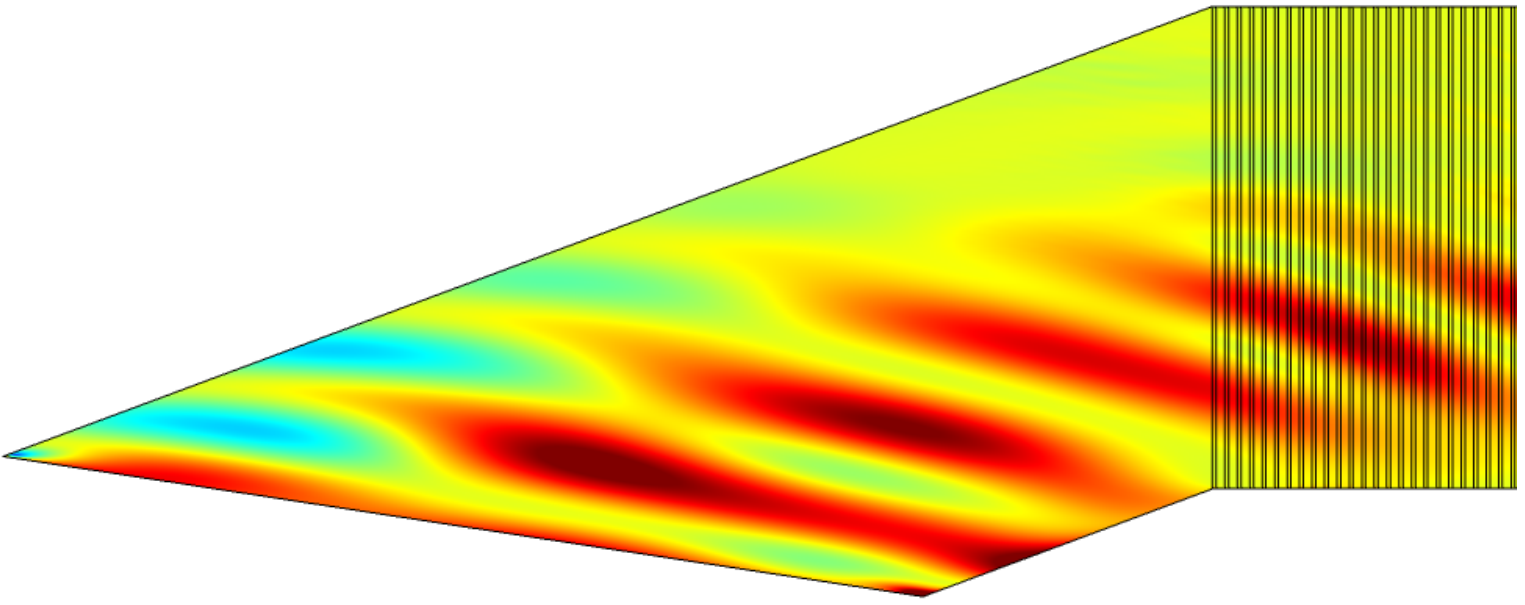}}  \caption{(a) Coalescence of three Bloch modes of the exemplary conservative laminate considered by \citet{FISHMAN2024JMPS}.  This EP3 is shown in a cut in the $(f,k_\mathrm{B})$ diagram for normalized vertical wavenumber $k_{2}h=0.15$ about 4.15\,kHz. Re$\,k_\mathrm{B}$ and Im$\,k_\mathrm{B}$ are shown in solid and dashed curves, respectively. The real and imaginary parts of a certain branch are plotted using the same color. (b) Full-wave finite element simulations of the axial energy flux in the pertinent scattering problem near the EP, giving rise to axially frozen modes with unity transmittance (see also Supplementary video 2).}

	{\small{}{}\label{fig:epthree}}{\small\par}
\end{figure}

In a subsequent work\cite{FISHMAN2024JMPS}, Shmuel's group designed the coalescence of three Bloch modes (EP3) by replacing one of isotropic materials in the unit cell with anisotropic one.  This non-resonant higher degeneracy further expands the toolbox for elastic wave shaping by forming waves with zero axial group velocity and finite transmittance (known as 'axially frozen modes'\cite{Figotin2003pre}), which can even reach unity (Fig.\ \ref{fig:epthree}b and supplementary video 2). Notably, these modes, which were previously discovered in 3D dielectric laminates\cite{Figotin2003pre}, are now accessible in simpler, planar settings in elastodynamics, thanks to its distinct tensorial richness.

\floatsetup[figure]{style=plain}
\captionsetup[subfloat]{position=top,labelformat=parens}

Additional useful phenomena are accessible by combining non-resonant EPs with gain and loss. For example, such combination enables the encirclement of non-resonant EPs in a suitable parameter space, a process that is known to be asymmetrically non-adiabatic\cite{Heiss2016yt,Doppler2016nature,Xu2016nature}. This means that the physical system will follow its instantaneous eigenmode when subjected to a loop in the parameter space, only along one of the two possible orientations of the loop. The solid (counterclockwise) and dashed (clockwise) curves in Fig.\ \ref{fig:eptwo}b demonstrate such adiabatic and non-adiabatic state evolution, respectively. \citet{elbaz2022jphysd} carried out such an encirclement around non-resonant EP2 of forward and backward longitudinal Bloch waves in elastic laminates, using spatial modulation of gain and loss. They discovered that the starting point of the loop governs several unusual features. For example, the laminate may act as a source or a sink of energy, exhibit reflectance greater than unity, and  accommodate spatial asymmetry in the energy flow with respect to the incidence direction, depending on that starting point. Encircling EPs provides another unexpected benefit in the form of a highly efficient algorithm for sorting eigenvalue bands, as observed by \citet{lu2018level}, who exploited EPs to distinguish real crossings from level repulsion zones in the real phononic spectrum.

\section{\label{sec:future directions}Future directions}
A future avenue that has been thoroughly explored in acoustics and photonics concerns non-Hermitian topology \cite{zhang2023second}. Non-Hermitian topology refers to the study of topological properties in systems that do not satisfy the Hermitian symmetry. In other words, non-Hermitian topology explores the emergence of topological phenomena in systems containing lossy and amplifying components. One of the key concepts in this field is the non-Hermitian skin effect, which refers to the localization of eigenstates at the boundary of a system.  As we know, the Bloch wave description is a powerful concept used to describe periodic states and resonances.  However, in non-Hermitian systems, where gain and loss mechanisms are present, the Bloch wave description encounters challenges. Using planar mechanical metamaterials, several numerical\cite{chen2019mechanical} and experimental \cite{wang2022non,wang2023non} works demonstrated how the eigenstates become localized at the boundary rather than uniformly distributed throughout the bulk. Targeting elastodynamic non-Hermitian bulk states at interfaces beyond the plane remain an intriguing area of research to pursue. Other exciting directions include combining lattice symmetry and non-Hermiticity and other phenomena that are elusive in elastodynamics such as non-Hermitian Weyl exceptional rings, non-Hermitian higher-order topology with vibrating complex corner states, and non-Abelian permutations\cite{zhang2023second} of mechanical states.

A completely uncharted direction is the breaking of Hermitian symmetry in materials that display electromomentum couplings\cite{PernasSalomon2019JMPS,pernassalomn2020prapplied,rps20201wm,muhafra2021,KOSTA2022ijss,Muhafra2023PRApplied}. The electromomentum couplings were theoretically discovered by \citet{PernasSalomon2019JMPS} using a generalization of Willis' dynamic homogenization method\cite{willis2011effective} to composites made of constituents that mechanically respond to non-mechanical stimuli. For the case of piezoelectric constituents, the method revealed that the electric displacement field ($\mathbf{D}$) constitutively depends on velocity and that the macroscopic linear momentum constitutively depends on the electric field ($-\nabla\phi$). These two couplings are captured in terms of two second-order tensors ($\mathbf{W}$ and $\hat{\mathbf{W}}$) called the electromomentum tensors, in direct analogy with the Willis tensors. These coupling tensors capture macroscale effects resulting from spatial symmetry-breaking on the microscale and nonlocal interactions\cite{PernasSalomon2019JMPS,pernassalomn2020prapplied,rps20201wm}. In symbolic matrix notation, the spatially local limit of these constitutive equations is
\begin{eqnarray}
\left(\begin{array}{c}
\boldsymbol{\sigma}\\
\mathbf{D}\\
\mathbf{p}
\end{array}\right)=\left(\begin{array}{ccc}
\mathbf{C} & \hat{\mathbf{B}} & \hat{\mathbf{S}}\\
\mathbf{B} & -\mathbf{A} & \mathbf{W}\\
\mathbf{S} & \hat{\mathbf{W}} & \boldsymbol{\rho}
\end{array}\right)\left(\begin{array}{c}
\grad\disp\\
\grad\phi\\
\dot{\disp}
\end{array}\right),\label{Eff-constR-final}
\end{eqnarray} where $\mathbf{A}$ is the permittivity tensor, $\mathbf{B}$ and $ \hat{\mathbf{B}}$ capture the direct- and inverse piezoelectric effect; collectively, these equations take a trianisotropic form\cite{shmuel2022eto}. Note that the constituents of such composites are piezoelectric and do not display Willis nor electromomentum coupling.  To date, studies\cite{Danawe2023APL, wallen2022polarizability,Lee2023JASA,HUYNH2023EML} have been restricted to materials with electromomentum tensors that are Hermitian adjoints\cite{pernassalomn2020prapplied}, which implies that the local electromomentum tensors satisfy $\hat{W}_{ij}=W_{ji}$. However, considering the works listed in Sec.\ \ref{sec:state of the art} which exploit piezoelectricity to generate material gain, this platform naturally lends itself to non-Hermitian physics, the simplest of which is of odd electromomentum tensors ($W_{ij}=-\hat{W}_{ji}$, right circle in Fig.\ \ref{fig:odd}).   The mplications of breaking this symmetry on elastic waves and how to realize designs with odd electromomentum tensors are yet to be explored, and constitute excellent candidates to extend this approach for elastic wave control.


\section*{\label{sec:ach}acknowledgements} GS acknowledges funding by the European Union (ERC, EXCEPTIONAL, Project No.~101045494); and the Israel Science Foundation, funded by the Israel Academy of Sciences and Humanities (Grant no.~2061/20). GLH acknowledges the Air Force Office of Scientific Research under Grant No.~AF9550-20-1-0279 with Program Manager Dr.~Byung-Lip (Les) Lee. MRH acknowledges support from Office of Naval Research under Award No.~N00014-23-1-2660. AS acknowledges the support from NSF grant No.~2219203 to the Illinois Institute of Technology.

\appendix 

\section{\label{app:non-herm}Non-Hermitian operators in elastodynamics}

  The dynamics of a solid occupying some domain $\Omega$ and subjected to the body force density $\mathbf{f}$ is governed by the balance of linear momentum $\sigma_{ij,j}+f_i=\dot{p}_i$, where  $\boldsymbol{\sigma}$ and $\mathbf{p}$ are the Cauchy stress tensor and linear momentum density, respectively.  In standard linear lossless and passive solids, the stress and linear momentum density constitutively depend on the displacement gradient $\nabla\mathbf{u}$ and velocity $\dot{\mathbf{u}}$ via the relations $\sigma_{ij}=C_{ijkl}u_{k,l}$ and $p_i=\rho \dot{u}_i$, where $\bm{C}$ is the (possibly anisotropic) elasticity tensor with major symmetry (owing to energy conservation\cite{ogden97book}) and $\rho$ is the mass density. When $\Omega$ is infinite (in 1,2, or 3 dimensions) and homogeneous, the equations simplify to $C_{ijkl}u_{k,lj}+f_i=\rho\ddot{u}_i$ and admit homogeneous solutions of the form $\bm{u}=\tilde{\bm{u}}\exp(ikn_jx_j-\omega t)$ where  $\omega$ is the frequency and $\boldsymbol{k}=kn_j\bm{e}_j$ is the wavevector with wavenumber $k$ and $\bm{e}_j$ denotes the direction of propagation. On substituting this ansatz in the equations of motion, one can formulate various generalized parametric eigenvalue problems which can be represented by linear operators. The simplest of these is $\bm{A}(\bm{k})\tilde{\bm{u}}=\lambda\bm{B}\tilde{\bm{u}}$ where in index notation the operators are $\lambda=-\omega^2$ and $A_{ij}=-k^2C_{ijkl}n_jn_l,\,B_{ij}=\rho\delta_{ij}$. In the absence of loss and gain and if $\bm{k}$ is real, this represents a Hermitian (or self-adjoint) problem in the following sense: if an inner product is defined as $\inner{\tilde{\bm{u}}}{\tilde{\bm{v}}}=\tilde{u}_i\tilde{v}_i^*$ then $\inner{\bm{A}\tilde{\bm{u}}}{\tilde{\bm{v}}}=\inner{\tilde{\bm{u}}}{\bm{A}\tilde{\bm{v}}}$ where we note that $\bm{B}$ is clearly Hermitian in this sense. This gives rise to exclusively real $\omega$ solutions for the eigenvalue problem considered. In standard elastodynamics literature\cite{auld1973acoustic,nayfeh1995wave,carcione2001wave}, this standard eigenvalue problem for $\lambda=\omega^2$ is known as the Christoffel matrix equation $[k^2\mathsf{\Gamma}]\{\mathsf{u}\}=\rho \omega^2\{\mathsf{u}\}$ where the components of the acoustic (or Christoffel) matrix are $\Gamma_{il}=C_{ijkl}n_j n_l$. The major symmetry of $\mathbf{C}$ implies that the real square matrix $\mathsf{\Gamma}$ is symmetric, hence has real eigenvalues $\omega^2$ and orthogonal eigenvectors. As discussed in Secs.\ \ref{sec:background} and \ref{sec:state of the art}, most works on non-Hermitian physics focus on this type of problems in which the eigenvalue is the natural frequency (squared), and use gain (or loss) to break Hermitian symmetry. This temporal energy transfer is modeled by complex $\bm{C}$ (or $\rho$), in which case $\bm{A}$ (equivalently, $\mathsf{\Gamma}$) is no longer Hermitian and therefore the solutions for $\omega$ are complex-valued in general even for real-valued $\bm{k}$.

  In a dual class of problems, the balance of linear momentum is formulated as an eigenvalue problem  for the wavenumber and the frequency is a prescribed real quantity. Mathematically, $\bm{A}$ is considered as a linear operator that depends upon a set of complex parameters $(\bm{k},\omega)$ and in this formulation the possibly complex wavevectors $\bm{k}$ are viewed as those resulting from the restriction of $\bm{A}(\omega);\omega\in\mathbb{C}$ to $\bm{A}(\omega);\, \omega\in\mathbb{R}$. 
  Physically, this formulation models structured conservative media subjected to sustained driving sources and open boundaries, such that there is no temporal energy loss or gain (real frequencies); the complex nature of the wavenumbers models spatially evanescent modes that accommodate boundary (or continuity) conditions at material interfaces\cite{graff1975wave,Srivastava2017PRA}, and implies that the corresponding operator is non-Hermitian even in the absence of gain or loss\cite{mokhtari2019properties}. It is sufficient to illustrate this by way of a simple planar example. Consider a medium that is $h$-periodic along $x_1$ such that the Bloch ansatz applies: $\bm{u}=\bm{\tilde{u}}(x_1)\exp(ikn_jx_j-\omega t)$ where $\bm{\tilde{u}},\, \rho,\, \bm{C}$ are $h$-periodic. Before we discuss the non-Hermitian problem, it is imperative to recall that the mainstay of phononic crystals research was the classical band structure $\omega(\bm{k})$ emanating from the eigenvalue problem  $\bm{A}(\bm{k})\tilde{\bm{u}}=\lambda\bm{B}\tilde{\bm{u}}$ with $\lambda=-\omega^2$ at real $\mathbf{k}$, where now $\bm{A}$ and $\bm{B}$ are linear differential operators. The restriction $\mathbf{k}\in\mathbb{R}$ applies to infinite periodic media, whose translational invariance prohibits complex-valued $\mathbf{k}$, and results with real $\omega$ solutions in the absence of gain or loss\cite{mokhtari2019properties}. This Hermitian problem has orthogonal eigenmodes and complete basis with respect to the inner product  $\inner{\tilde{\bm{u}}}{\tilde{\bm{v}}}=\int_h\tilde{u}_i\tilde{v}_i^*dh$. 
  
  The operator loses its Hermitian nature once $\bm{k}$ is extended to the complex plane. As we discussed above, complex $\bm{k}$ has a clear physical meaning of evanescent waves in interface problems. For example, complex $\bm{k}$ captures the modes that would evolve in the periodic medium if it is truncated at some position $x_1$, and subjected to incoming waves from a homogeneous half-space occupying the volume $x_1<0$. Thus, we further specialize our problem to a scenario where  $k_2$ is real and prescribed (physically, by defining the incident wave), to obtain the  generalized eigenvalue problem $\bm{A}(\omega,k_2)\tilde{\bm{u}}=k_1\bm{B}(\omega,k_2)\tilde{\bm{u}}$. As discussed in Sec.\ \ref{sec:state of the art}, \citet{Lustig2019} highlighted the non-Hermitian nature of this classical problem, identifying EPs inside the Brillouin zone (Fig.\ \ref{fig:eptwo}a) and the Riemann sheets around them in the complex wavevector space (Fig.\ \ref{fig:eptwo}b). \citet{mokhtari2020scattering} later highlighted another aspect of the non-Hermitian nature, namely, the coalescence of the spatial eigenmodes at these EPs. Again, this problem is non-Hermitian in spite of the fact that the media is conservative (real symmetric elasticity tensor and real mass density). We reemphasize that of all the classical systems, EPs between forward waves in such simple settings (planar layered medium made of isotropic conservative materials)  can form only in elastodynamics\cite{Lustig2019}, owing to its distinct co-existence of transverse and longitudinal wave polarizations. Recent works in elastodynamics\cite{khodavirdi2022scattering,khodavirdi2023atomistic} and acoustic-elastodynamic interactions leading to scattering\cite{zhou2024underwater} address such problems using open system formalism which leads to a non-Hermitian effective Hamiltonian. 

We complete this Appendix recalling that Chauchy-elastic materials, Willis materials and electromomentum materials offer a richer platform to break Hermitian symmetry, even in the first type of eigenvalues problems for $\omega^2$ in infinite homogeneous media. If the major symmetry of $\mathbf{C}$ is broken, then the operator for  $\omega$ is no longer Hermitian and complex solutions will form. 
Similarly, if a Willis material is designed such that one of the symmetries $C_{ijkl}=C_{klij},\ \rho_{ij}=\rho_{ji},\  \hat{S}_{ijk}=S_{kij}$ is broken, then the operator in the first type of problems is non-Hermitian, giving rise to complex $\omega$ solutions. In direct analogy, electromomentum couplings constitute an additional degree of freedom to break Hermitian symmetry: if the microstructure is designed such that $W_{ij}\neq\hat{W}_{ji}$, then the operator in the first eigenvalue problem is non-Hermitian and its eigen-frequencies are complex.

\bibliography{aipsamp,bibtexfiletot,library-as}

\providecommand{\noopsort}[1]{}\providecommand{\singleletter}[1]{#1}%
\begin{thebibliography}{117}%
\makeatletter
\providecommand \@ifxundefined [1]{%
 \@ifx{#1\undefined}
}%
\providecommand \@ifnum [1]{%
 \ifnum #1\expandafter \@firstoftwo
 \else \expandafter \@secondoftwo
 \fi
}%
\providecommand \@ifx [1]{%
 \ifx #1\expandafter \@firstoftwo
 \else \expandafter \@secondoftwo
 \fi
}%
\providecommand \natexlab [1]{#1}%
\providecommand \enquote  [1]{``#1''}%
\providecommand \bibnamefont  [1]{#1}%
\providecommand \bibfnamefont [1]{#1}%
\providecommand \citenamefont [1]{#1}%
\providecommand \href@noop [0]{\@secondoftwo}%
\providecommand \href [0]{\begingroup \@sanitize@url \@href}%
\providecommand \@href[1]{\@@startlink{#1}\@@href}%
\providecommand \@@href[1]{\endgroup#1\@@endlink}%
\providecommand \@sanitize@url [0]{\catcode `\\12\catcode `\$12\catcode
  `\&12\catcode `\#12\catcode `\^12\catcode `\_12\catcode `\%12\relax}%
\providecommand \@@startlink[1]{}%
\providecommand \@@endlink[0]{}%
\providecommand \url  [0]{\begingroup\@sanitize@url \@url }%
\providecommand \@url [1]{\endgroup\@href {#1}{\urlprefix }}%
\providecommand \urlprefix  [0]{URL }%
\providecommand \Eprint [0]{\href }%
\providecommand \doibase [0]{http://dx.doi.org/}%
\providecommand \selectlanguage [0]{\@gobble}%
\providecommand \bibinfo  [0]{\@secondoftwo}%
\providecommand \bibfield  [0]{\@secondoftwo}%
\providecommand \translation [1]{[#1]}%
\providecommand \BibitemOpen [0]{}%
\providecommand \bibitemStop [0]{}%
\providecommand \bibitemNoStop [0]{.\EOS\space}%
\providecommand \EOS [0]{\spacefactor3000\relax}%
\providecommand \BibitemShut  [1]{\csname bibitem#1\endcsname}%
\let\auto@bib@innerbib\@empty
\bibitem [{\citenamefont {Cummer}, \citenamefont {Christensen},\ and\
  \citenamefont {Al{\`u}}(2016)}]{cummer2016controlling}%
  \BibitemOpen
  \bibfield  {author} {\bibinfo {author} {\bibfnamefont {S.~A.}\ \bibnamefont
  {Cummer}}, \bibinfo {author} {\bibfnamefont {J.}~\bibnamefont {Christensen}},
  \ and\ \bibinfo {author} {\bibfnamefont {A.}~\bibnamefont {Al{\`u}}},\
  }\bibfield  {title} {\enquote {\bibinfo {title} {Controlling sound with
  acoustic metamaterials},}\ }\href@noop {} {\bibfield  {journal} {\bibinfo
  {journal} {Nature Reviews Materials}\ }\textbf {\bibinfo {volume} {1}},\
  \bibinfo {pages} {1--13} (\bibinfo {year} {2016})}\BibitemShut {NoStop}%
\bibitem [{\citenamefont {Haberman}\ and\ \citenamefont
  {Guild}(2016)}]{haberman2016acoustic}%
  \BibitemOpen
  \bibfield  {author} {\bibinfo {author} {\bibfnamefont {M.~R.}\ \bibnamefont
  {Haberman}}\ and\ \bibinfo {author} {\bibfnamefont {M.~D.}\ \bibnamefont
  {Guild}},\ }\bibfield  {title} {\enquote {\bibinfo {title} {Acoustic
  metamaterials},}\ }\href@noop {} {\bibfield  {journal} {\bibinfo  {journal}
  {Physics Today}\ }\textbf {\bibinfo {volume} {69}},\ \bibinfo {pages}
  {42--48} (\bibinfo {year} {2016})}\BibitemShut {NoStop}%
\bibitem [{\citenamefont {Kochmann}\ and\ \citenamefont
  {Bertoldi}(2017)}]{kochmann2017exploiting}%
  \BibitemOpen
  \bibfield  {author} {\bibinfo {author} {\bibfnamefont {D.~M.}\ \bibnamefont
  {Kochmann}}\ and\ \bibinfo {author} {\bibfnamefont {K.}~\bibnamefont
  {Bertoldi}},\ }\bibfield  {title} {\enquote {\bibinfo {title} {Exploiting
  microstructural instabilities in solids and structures: from metamaterials to
  structural transitions},}\ }\href@noop {} {\bibfield  {journal} {\bibinfo
  {journal} {Applied mechanics reviews}\ }\textbf {\bibinfo {volume} {69}},\
  \bibinfo {pages} {050801} (\bibinfo {year} {2017})}\BibitemShut {NoStop}%
\bibitem [{\citenamefont {Kadic}\ \emph {et~al.}(2019)\citenamefont {Kadic},
  \citenamefont {Milton}, \citenamefont {van Hecke},\ and\ \citenamefont
  {Wegener}}]{Kadi2019nrp}%
  \BibitemOpen
  \bibfield  {author} {\bibinfo {author} {\bibfnamefont {M.}~\bibnamefont
  {Kadic}}, \bibinfo {author} {\bibfnamefont {G.~W.}\ \bibnamefont {Milton}},
  \bibinfo {author} {\bibfnamefont {M.}~\bibnamefont {van Hecke}}, \ and\
  \bibinfo {author} {\bibfnamefont {M.}~\bibnamefont {Wegener}},\ }\bibfield
  {title} {\enquote {\bibinfo {title} {3d metamaterials},}\ }\href {\doibase
  10.1038/s42254-018-0018-y} {\bibfield  {journal} {\bibinfo  {journal} {Nature
  Reviews Physics}\ }\textbf {\bibinfo {volume} {1}},\ \bibinfo {pages}
  {198--210} (\bibinfo {year} {2019})}\BibitemShut {NoStop}%
\bibitem [{\citenamefont {Bigoni}\ \emph {et~al.}(2013)\citenamefont {Bigoni},
  \citenamefont {Guenneau}, \citenamefont {Movchan},\ and\ \citenamefont
  {Brun}}]{Bigoni2013PRB}%
  \BibitemOpen
  \bibfield  {author} {\bibinfo {author} {\bibfnamefont {D.}~\bibnamefont
  {Bigoni}}, \bibinfo {author} {\bibfnamefont {S.}~\bibnamefont {Guenneau}},
  \bibinfo {author} {\bibfnamefont {A.~B.}\ \bibnamefont {Movchan}}, \ and\
  \bibinfo {author} {\bibfnamefont {M.}~\bibnamefont {Brun}},\ }\bibfield
  {title} {\enquote {\bibinfo {title} {{Elastic metamaterials with inertial
  locally resonant structures: Application to lensing and localization}},}\
  }\href {\doibase 10.1103/PhysRevB.87.174303} {\bibfield  {journal} {\bibinfo
  {journal} {Phys. Rev. B}\ }\textbf {\bibinfo {volume} {87}},\ \bibinfo
  {pages} {174303} (\bibinfo {year} {2013})}\BibitemShut {NoStop}%
\bibitem [{\citenamefont {Christensen}\ \emph {et~al.}(2015)\citenamefont
  {Christensen}, \citenamefont {Kadic}, \citenamefont {Kraft},\ and\
  \citenamefont {Wegener}}]{Christensen2015MRCComunications}%
  \BibitemOpen
  \bibfield  {author} {\bibinfo {author} {\bibfnamefont {J.}~\bibnamefont
  {Christensen}}, \bibinfo {author} {\bibfnamefont {M.}~\bibnamefont {Kadic}},
  \bibinfo {author} {\bibfnamefont {O.}~\bibnamefont {Kraft}}, \ and\ \bibinfo
  {author} {\bibfnamefont {M.}~\bibnamefont {Wegener}},\ }\bibfield  {title}
  {\enquote {\bibinfo {title} {Vibrant times for mechanical metamaterials},}\
  }\href {\doibase 10.1557/mrc.2015.51} {\bibfield  {journal} {\bibinfo
  {journal} {MRS Communications}\ }\textbf {\bibinfo {volume} {5}},\ \bibinfo
  {pages} {453--462} (\bibinfo {year} {2015})}\BibitemShut {NoStop}%
\bibitem [{\citenamefont {Celli}\ and\ \citenamefont
  {Gonella}(2015)}]{gonella15}%
  \BibitemOpen
  \bibfield  {author} {\bibinfo {author} {\bibfnamefont {P.}~\bibnamefont
  {Celli}}\ and\ \bibinfo {author} {\bibfnamefont {S.}~\bibnamefont
  {Gonella}},\ }\bibfield  {title} {\enquote {\bibinfo {title} {{Manipulating
  waves with LEGO{\textregistered}bricks: A versatile experimental platform for
  metamaterial architectures}},}\ }\href {\doibase
  http://dx.doi.org/10.1063/1.4929566} {\bibfield  {journal} {\bibinfo
  {journal} {Appl. Phys. Lett.}\ }\textbf {\bibinfo {volume} {107}} (\bibinfo
  {year} {2015}),\ http://dx.doi.org/10.1063/1.4929566}\BibitemShut {NoStop}%
\bibitem [{\citenamefont {Nadkarni}, \citenamefont {Daraio},\ and\
  \citenamefont {Kochmann}(2014)}]{kochmann14pre}%
  \BibitemOpen
  \bibfield  {author} {\bibinfo {author} {\bibfnamefont {N.}~\bibnamefont
  {Nadkarni}}, \bibinfo {author} {\bibfnamefont {C.}~\bibnamefont {Daraio}}, \
  and\ \bibinfo {author} {\bibfnamefont {D.~M.}\ \bibnamefont {Kochmann}},\
  }\bibfield  {title} {\enquote {\bibinfo {title} {{Dynamics of periodic
  mechanical structures containing bistable elastic elements: From elastic to
  solitary wave propagation}},}\ }\href {\doibase 10.1103/PhysRevE.90.023204}
  {\bibfield  {journal} {\bibinfo  {journal} {Phys. Rev. E}\ }\textbf {\bibinfo
  {volume} {90}},\ \bibinfo {pages} {23204} (\bibinfo {year}
  {2014})}\BibitemShut {NoStop}%
\bibitem [{\citenamefont {Parnell}, \citenamefont {Norris},\ and\ \citenamefont
  {Shearer}(2012)}]{parnell2012employing}%
  \BibitemOpen
  \bibfield  {author} {\bibinfo {author} {\bibfnamefont {W.~J.}\ \bibnamefont
  {Parnell}}, \bibinfo {author} {\bibfnamefont {A.~N.}\ \bibnamefont {Norris}},
  \ and\ \bibinfo {author} {\bibfnamefont {T.}~\bibnamefont {Shearer}},\
  }\bibfield  {title} {\enquote {\bibinfo {title} {{Employing pre-stress to
  generate finite cloaks for antiplane elastic waves}},}\ }\href@noop {}
  {\bibfield  {journal} {\bibinfo  {journal} {Applied Physics Letters}\
  }\textbf {\bibinfo {volume} {100}},\ \bibinfo {pages} {171907} (\bibinfo
  {year} {2012})}\BibitemShut {NoStop}%
\bibitem [{\citenamefont {Bender}\ and\ \citenamefont
  {Boettcher}(1998)}]{bender1998real}%
  \BibitemOpen
  \bibfield  {author} {\bibinfo {author} {\bibfnamefont {C.~M.}\ \bibnamefont
  {Bender}}\ and\ \bibinfo {author} {\bibfnamefont {S.}~\bibnamefont
  {Boettcher}},\ }\bibfield  {title} {\enquote {\bibinfo {title} {{Real spectra
  in non-Hermitian Hamiltonians having P T symmetry}},}\ }\href@noop {}
  {\bibfield  {journal} {\bibinfo  {journal} {Physical Review Letters}\
  }\textbf {\bibinfo {volume} {80}},\ \bibinfo {pages} {5243} (\bibinfo {year}
  {1998})}\BibitemShut {NoStop}%
\bibitem [{\citenamefont {Mostafazadeh}(2010)}]{MOSTAFAZADEH2010ijgmp}%
  \BibitemOpen
  \bibfield  {author} {\bibinfo {author} {\bibfnamefont {A.}~\bibnamefont
  {Mostafazadeh}},\ }\bibfield  {title} {\enquote {\bibinfo {title}
  {Pseudo-hermitian representation of quantum mechanics},}\ }\href {\doibase
  10.1142/S0219887810004816} {\bibfield  {journal} {\bibinfo  {journal}
  {International Journal of Geometric Methods in Modern Physics}\ }\textbf
  {\bibinfo {volume} {07}},\ \bibinfo {pages} {1191--1306} (\bibinfo {year}
  {2010})},\ \Eprint
  {http://arxiv.org/abs/https://doi.org/10.1142/S0219887810004816}
  {https://doi.org/10.1142/S0219887810004816} \BibitemShut {NoStop}%
\bibitem [{\citenamefont {Moiseyev}(2011)}]{moiseyev2011book}%
  \BibitemOpen
  \bibfield  {author} {\bibinfo {author} {\bibfnamefont {N.}~\bibnamefont
  {Moiseyev}},\ }\href {\doibase 10.1017/CBO9780511976186} {\emph {\bibinfo
  {title} {{Non-Hermitian Quantum Mechanics}}}}\ (\bibinfo  {publisher}
  {Cambridge University Press},\ \bibinfo {year} {2011})\BibitemShut {NoStop}%
\bibitem [{\citenamefont {Feng}, \citenamefont {El-Ganainy},\ and\
  \citenamefont {Ge}(2017)}]{Feng2017np}%
  \BibitemOpen
  \bibfield  {author} {\bibinfo {author} {\bibfnamefont {L.}~\bibnamefont
  {Feng}}, \bibinfo {author} {\bibfnamefont {R.}~\bibnamefont {El-Ganainy}}, \
  and\ \bibinfo {author} {\bibfnamefont {L.}~\bibnamefont {Ge}},\ }\bibfield
  {title} {\enquote {\bibinfo {title} {Non-hermitian photonics based on
  parity--time symmetry},}\ }\href {\doibase 10.1038/s41566-017-0031-1}
  {\bibfield  {journal} {\bibinfo  {journal} {Nature Photonics}\ }\textbf
  {\bibinfo {volume} {11}},\ \bibinfo {pages} {752--762} (\bibinfo {year}
  {2017})}\BibitemShut {NoStop}%
\bibitem [{\citenamefont {El-Ganainy}\ \emph {et~al.}(2018)\citenamefont
  {El-Ganainy}, \citenamefont {Makris}, \citenamefont {Khajavikhan},
  \citenamefont {Musslimani}, \citenamefont {Rotter},\ and\ \citenamefont
  {Christodoulides}}]{El-Ganainy2018ys}%
  \BibitemOpen
  \bibfield  {author} {\bibinfo {author} {\bibfnamefont {R.}~\bibnamefont
  {El-Ganainy}}, \bibinfo {author} {\bibfnamefont {K.~G.}\ \bibnamefont
  {Makris}}, \bibinfo {author} {\bibfnamefont {M.}~\bibnamefont {Khajavikhan}},
  \bibinfo {author} {\bibfnamefont {Z.~H.}\ \bibnamefont {Musslimani}},
  \bibinfo {author} {\bibfnamefont {S.}~\bibnamefont {Rotter}}, \ and\ \bibinfo
  {author} {\bibfnamefont {D.~N.}\ \bibnamefont {Christodoulides}},\ }\bibfield
   {title} {\enquote {\bibinfo {title} {{Non-Hermitian physics and PT
  symmetry}},}\ }\href {https://doi.org/10.1038/nphys4323} {\bibfield
  {journal} {\bibinfo  {journal} {Nature Physics}\ }\textbf {\bibinfo {volume}
  {14}},\ \bibinfo {pages} {11 EP --} (\bibinfo {year} {2018})}\BibitemShut
  {NoStop}%
\bibitem [{\citenamefont {{\"O}zdemir}\ \emph {et~al.}(2019)\citenamefont
  {{\"O}zdemir}, \citenamefont {Rotter}, \citenamefont {Nori},\ and\
  \citenamefont {Yang}}]{Ozdemir2019cr}%
  \BibitemOpen
  \bibfield  {author} {\bibinfo {author} {\bibfnamefont {{\c S}.~K.}\
  \bibnamefont {{\"O}zdemir}}, \bibinfo {author} {\bibfnamefont
  {S.}~\bibnamefont {Rotter}}, \bibinfo {author} {\bibfnamefont
  {F.}~\bibnamefont {Nori}}, \ and\ \bibinfo {author} {\bibfnamefont
  {L.}~\bibnamefont {Yang}},\ }\bibfield  {title} {\enquote {\bibinfo {title}
  {Parity--time symmetry and exceptional points in photonics},}\ }\href
  {\doibase 10.1038/s41563-019-0304-9} {\bibfield  {journal} {\bibinfo
  {journal} {Nature Materials}\ }\textbf {\bibinfo {volume} {18}},\ \bibinfo
  {pages} {783--798} (\bibinfo {year} {2019})}\BibitemShut {NoStop}%
\bibitem [{\citenamefont {Miri}\ and\ \citenamefont
  {Al{\`u}}(2019)}]{Miri2019science}%
  \BibitemOpen
  \bibfield  {author} {\bibinfo {author} {\bibfnamefont {M.-A.}\ \bibnamefont
  {Miri}}\ and\ \bibinfo {author} {\bibfnamefont {A.}~\bibnamefont {Al{\`u}}},\
  }\bibfield  {title} {\enquote {\bibinfo {title} {Exceptional points in optics
  and photonics},}\ }\href {\doibase 10.1126/science.aar7709} {\bibfield
  {journal} {\bibinfo  {journal} {Science}\ }\textbf {\bibinfo {volume} {363}}
  (\bibinfo {year} {2019}),\ 10.1126/science.aar7709},\ \Eprint
  {http://arxiv.org/abs/https://science.sciencemag.org/content/363/6422/eaar7709.full.pdf}
  {https://science.sciencemag.org/content/363/6422/eaar7709.full.pdf}
  \BibitemShut {NoStop}%
\bibitem [{\citenamefont {Pile}(2017)}]{Pile:2017jk}%
  \BibitemOpen
  \bibfield  {author} {\bibinfo {author} {\bibfnamefont {I.~b. D. F.~P.}\
  \bibnamefont {Pile}},\ }\bibfield  {title} {\enquote {\bibinfo {title}
  {Gaining with loss},}\ }\href {\doibase 10.1038/s41566-017-0060-9} {\bibfield
   {journal} {\bibinfo  {journal} {Nature Photonics}\ }\textbf {\bibinfo
  {volume} {11}},\ \bibinfo {pages} {742--743} (\bibinfo {year}
  {2017})}\BibitemShut {NoStop}%
\bibitem [{\citenamefont {Midya}, \citenamefont {Zhao},\ and\ \citenamefont
  {Feng}(2018)}]{Midya2018natcom}%
  \BibitemOpen
  \bibfield  {author} {\bibinfo {author} {\bibfnamefont {B.}~\bibnamefont
  {Midya}}, \bibinfo {author} {\bibfnamefont {H.}~\bibnamefont {Zhao}}, \ and\
  \bibinfo {author} {\bibfnamefont {L.}~\bibnamefont {Feng}},\ }\bibfield
  {title} {\enquote {\bibinfo {title} {Non-hermitian photonics promises
  exceptional topology of light},}\ }\href {\doibase
  10.1038/s41467-018-05175-8} {\bibfield  {journal} {\bibinfo  {journal}
  {Nature Communications}\ }\textbf {\bibinfo {volume} {9}},\ \bibinfo {pages}
  {2674} (\bibinfo {year} {2018})}\BibitemShut {NoStop}%
\bibitem [{\citenamefont {El-Ganainy}\ \emph {et~al.}(2019)\citenamefont
  {El-Ganainy}, \citenamefont {Khajavikhan}, \citenamefont {Christodoulides},\
  and\ \citenamefont {Ozdemir}}]{El-Ganainy2019cp}%
  \BibitemOpen
  \bibfield  {author} {\bibinfo {author} {\bibfnamefont {R.}~\bibnamefont
  {El-Ganainy}}, \bibinfo {author} {\bibfnamefont {M.}~\bibnamefont
  {Khajavikhan}}, \bibinfo {author} {\bibfnamefont {D.~N.}\ \bibnamefont
  {Christodoulides}}, \ and\ \bibinfo {author} {\bibfnamefont {S.~K.}\
  \bibnamefont {Ozdemir}},\ }\bibfield  {title} {\enquote {\bibinfo {title}
  {The dawn of non-hermitian optics},}\ }\href {\doibase
  10.1038/s42005-019-0130-z} {\bibfield  {journal} {\bibinfo  {journal}
  {Communications Physics}\ }\textbf {\bibinfo {volume} {2}},\ \bibinfo {pages}
  {37} (\bibinfo {year} {2019})}\BibitemShut {NoStop}%
\bibitem [{\citenamefont {El-Ganainy}\ \emph {et~al.}(2007)\citenamefont
  {El-Ganainy}, \citenamefont {Makris}, \citenamefont {Christodoulides},\ and\
  \citenamefont {Musslimani}}]{el2007theory}%
  \BibitemOpen
  \bibfield  {author} {\bibinfo {author} {\bibfnamefont {R.}~\bibnamefont
  {El-Ganainy}}, \bibinfo {author} {\bibfnamefont {K.~G.}\ \bibnamefont
  {Makris}}, \bibinfo {author} {\bibfnamefont {D.~N.}\ \bibnamefont
  {Christodoulides}}, \ and\ \bibinfo {author} {\bibfnamefont {Z.~H.}\
  \bibnamefont {Musslimani}},\ }\bibfield  {title} {\enquote {\bibinfo {title}
  {{Theory of coupled optical PT-symmetric structures}},}\ }\href@noop {}
  {\bibfield  {journal} {\bibinfo  {journal} {Optics letters}\ }\textbf
  {\bibinfo {volume} {32}},\ \bibinfo {pages} {2632--2634} (\bibinfo {year}
  {2007})}\BibitemShut {NoStop}%
\bibitem [{\citenamefont {Klaiman}, \citenamefont {G{\"{u}}nther},\ and\
  \citenamefont {Moiseyev}(2008)}]{klaiman2008visualization}%
  \BibitemOpen
  \bibfield  {author} {\bibinfo {author} {\bibfnamefont {S.}~\bibnamefont
  {Klaiman}}, \bibinfo {author} {\bibfnamefont {U.}~\bibnamefont
  {G{\"{u}}nther}}, \ and\ \bibinfo {author} {\bibfnamefont {N.}~\bibnamefont
  {Moiseyev}},\ }\bibfield  {title} {\enquote {\bibinfo {title} {{Visualization
  of branch points in p t-symmetric waveguides}},}\ }\href@noop {} {\bibfield
  {journal} {\bibinfo  {journal} {Physical review letters}\ }\textbf {\bibinfo
  {volume} {101}},\ \bibinfo {pages} {80402} (\bibinfo {year}
  {2008})}\BibitemShut {NoStop}%
\bibitem [{\citenamefont {Guo}\ \emph {et~al.}(2009)\citenamefont {Guo},
  \citenamefont {Salamo}, \citenamefont {Duchesne}, \citenamefont {Morandotti},
  \citenamefont {Volatier-Ravat}, \citenamefont {Aimez}, \citenamefont
  {Siviloglou},\ and\ \citenamefont {Christodoulides}}]{guo2009observation}%
  \BibitemOpen
  \bibfield  {author} {\bibinfo {author} {\bibfnamefont {A.}~\bibnamefont
  {Guo}}, \bibinfo {author} {\bibfnamefont {G.~J.}\ \bibnamefont {Salamo}},
  \bibinfo {author} {\bibfnamefont {D.}~\bibnamefont {Duchesne}}, \bibinfo
  {author} {\bibfnamefont {R.}~\bibnamefont {Morandotti}}, \bibinfo {author}
  {\bibfnamefont {M.}~\bibnamefont {Volatier-Ravat}}, \bibinfo {author}
  {\bibfnamefont {V.}~\bibnamefont {Aimez}}, \bibinfo {author} {\bibfnamefont
  {G.~A.}\ \bibnamefont {Siviloglou}}, \ and\ \bibinfo {author} {\bibfnamefont
  {D.~N.}\ \bibnamefont {Christodoulides}},\ }\bibfield  {title} {\enquote
  {\bibinfo {title} {{Observation of P T-symmetry breaking in complex optical
  potentials}},}\ }\href@noop {} {\bibfield  {journal} {\bibinfo  {journal}
  {Physical Review Letters}\ }\textbf {\bibinfo {volume} {103}},\ \bibinfo
  {pages} {93902} (\bibinfo {year} {2009})}\BibitemShut {NoStop}%
\bibitem [{\citenamefont {R{\"{u}}ter}\ \emph {et~al.}(2010)\citenamefont
  {R{\"{u}}ter}, \citenamefont {Makris}, \citenamefont {El-Ganainy},
  \citenamefont {Christodoulides}, \citenamefont {Segev},\ and\ \citenamefont
  {Kip}}]{ruter2010observation}%
  \BibitemOpen
  \bibfield  {author} {\bibinfo {author} {\bibfnamefont {C.~E.}\ \bibnamefont
  {R{\"{u}}ter}}, \bibinfo {author} {\bibfnamefont {K.~G.}\ \bibnamefont
  {Makris}}, \bibinfo {author} {\bibfnamefont {R.}~\bibnamefont {El-Ganainy}},
  \bibinfo {author} {\bibfnamefont {D.~N.}\ \bibnamefont {Christodoulides}},
  \bibinfo {author} {\bibfnamefont {M.}~\bibnamefont {Segev}}, \ and\ \bibinfo
  {author} {\bibfnamefont {D.}~\bibnamefont {Kip}},\ }\bibfield  {title}
  {\enquote {\bibinfo {title} {{Observation of parity--time symmetry in
  optics}},}\ }\href@noop {} {\bibfield  {journal} {\bibinfo  {journal} {Nature
  physics}\ }\textbf {\bibinfo {volume} {6}},\ \bibinfo {pages} {192} (\bibinfo
  {year} {2010})}\BibitemShut {NoStop}%
\bibitem [{\citenamefont {Longhi}(2017{\natexlab{a}})}]{Longhi2017}%
  \BibitemOpen
  \bibfield  {author} {\bibinfo {author} {\bibfnamefont {S.}~\bibnamefont
  {Longhi}},\ }\bibfield  {title} {\enquote {\bibinfo {title} {Parity-time
  symmetry meets photonics: A new twist in non-hermitian optics},}\ }\href
  {\doibase 10.1209/0295-5075/120/64001} {\bibfield  {journal} {\bibinfo
  {journal} {{EPL} (Europhysics Letters)}\ }\textbf {\bibinfo {volume} {120}},\
  \bibinfo {pages} {64001} (\bibinfo {year} {2017}{\natexlab{a}})}\BibitemShut
  {NoStop}%
\bibitem [{\citenamefont {Goldzak}, \citenamefont {Mailybaev},\ and\
  \citenamefont {Moiseyev}(2018)}]{Goldzak2018PRL}%
  \BibitemOpen
  \bibfield  {author} {\bibinfo {author} {\bibfnamefont {T.}~\bibnamefont
  {Goldzak}}, \bibinfo {author} {\bibfnamefont {A.~A.}\ \bibnamefont
  {Mailybaev}}, \ and\ \bibinfo {author} {\bibfnamefont {N.}~\bibnamefont
  {Moiseyev}},\ }\bibfield  {title} {\enquote {\bibinfo {title} {{Light Stops
  at Exceptional Points}},}\ }\href {\doibase 10.1103/PhysRevLett.120.013901}
  {\bibfield  {journal} {\bibinfo  {journal} {Phys. Rev. Lett.}\ }\textbf
  {\bibinfo {volume} {120}},\ \bibinfo {pages} {13901} (\bibinfo {year}
  {2018})}\BibitemShut {NoStop}%
\bibitem [{\citenamefont {Zhao}\ and\ \citenamefont
  {Feng}(2018)}]{Zhao2018nsr}%
  \BibitemOpen
  \bibfield  {author} {\bibinfo {author} {\bibfnamefont {H.}~\bibnamefont
  {Zhao}}\ and\ \bibinfo {author} {\bibfnamefont {L.}~\bibnamefont {Feng}},\
  }\bibfield  {title} {\enquote {\bibinfo {title} {{Parity--time symmetric
  photonics}},}\ }\href {\doibase 10.1093/nsr/nwy011} {\bibfield  {journal}
  {\bibinfo  {journal} {National Science Review}\ }\textbf {\bibinfo {volume}
  {5}},\ \bibinfo {pages} {183--199} (\bibinfo {year} {2018})},\ \Eprint
  {http://arxiv.org/abs/http://oup.prod.sis.lan/nsr/article-pdf/5/2/183/24511849/nwy011.pdf}
  {http://oup.prod.sis.lan/nsr/article-pdf/5/2/183/24511849/nwy011.pdf}
  \BibitemShut {NoStop}%
\bibitem [{\citenamefont {Zhu}\ \emph {et~al.}(2014)\citenamefont {Zhu},
  \citenamefont {Ramezani}, \citenamefont {Shi}, \citenamefont {Zhu},\ and\
  \citenamefont {Zhang}}]{zhu2014p}%
  \BibitemOpen
  \bibfield  {author} {\bibinfo {author} {\bibfnamefont {X.}~\bibnamefont
  {Zhu}}, \bibinfo {author} {\bibfnamefont {H.}~\bibnamefont {Ramezani}},
  \bibinfo {author} {\bibfnamefont {C.}~\bibnamefont {Shi}}, \bibinfo {author}
  {\bibfnamefont {J.}~\bibnamefont {Zhu}}, \ and\ \bibinfo {author}
  {\bibfnamefont {X.}~\bibnamefont {Zhang}},\ }\bibfield  {title} {\enquote
  {\bibinfo {title} {{P t-symmetric acoustics}},}\ }\href@noop {} {\bibfield
  {journal} {\bibinfo  {journal} {Physical Review X}\ }\textbf {\bibinfo
  {volume} {4}},\ \bibinfo {pages} {31042} (\bibinfo {year}
  {2014})}\BibitemShut {NoStop}%
\bibitem [{\citenamefont {Fleury}, \citenamefont {Sounas},\ and\ \citenamefont
  {Al{\`{u}}}(2015)}]{Fleury2015}%
  \BibitemOpen
  \bibfield  {author} {\bibinfo {author} {\bibfnamefont {R.}~\bibnamefont
  {Fleury}}, \bibinfo {author} {\bibfnamefont {D.}~\bibnamefont {Sounas}}, \
  and\ \bibinfo {author} {\bibfnamefont {A.}~\bibnamefont {Al{\`{u}}}},\
  }\bibfield  {title} {\enquote {\bibinfo {title} {{An invisible acoustic
  sensor based on parity-time symmetry}},}\ }\href {\doibase
  10.1038/ncomms6905} {\bibfield  {journal} {\bibinfo  {journal} {Nature
  Communications}\ }\textbf {\bibinfo {volume} {6}},\ \bibinfo {pages} {5905}
  (\bibinfo {year} {2015})}\BibitemShut {NoStop}%
\bibitem [{\citenamefont {Cummer}, \citenamefont {Christensen},\ and\
  \citenamefont {Al{\`{u}}}(2016)}]{Cummer2016}%
  \BibitemOpen
  \bibfield  {author} {\bibinfo {author} {\bibfnamefont {S.~A.}\ \bibnamefont
  {Cummer}}, \bibinfo {author} {\bibfnamefont {J.}~\bibnamefont {Christensen}},
  \ and\ \bibinfo {author} {\bibfnamefont {A.}~\bibnamefont {Al{\`{u}}}},\
  }\href {\doibase 10.1038/natrevmats.2016.1} {\enquote {\bibinfo {title}
  {{Controlling sound with acoustic metamaterials}},}\ } (\bibinfo {year}
  {2016})\BibitemShut {NoStop}%
\bibitem [{\citenamefont {Shi}\ \emph {et~al.}(2016)\citenamefont {Shi},
  \citenamefont {Dubois}, \citenamefont {Chen}, \citenamefont {Cheng},
  \citenamefont {Ramezani}, \citenamefont {Wang},\ and\ \citenamefont
  {Zhang}}]{shi2016accessing}%
  \BibitemOpen
  \bibfield  {author} {\bibinfo {author} {\bibfnamefont {C.}~\bibnamefont
  {Shi}}, \bibinfo {author} {\bibfnamefont {M.}~\bibnamefont {Dubois}},
  \bibinfo {author} {\bibfnamefont {Y.}~\bibnamefont {Chen}}, \bibinfo {author}
  {\bibfnamefont {L.}~\bibnamefont {Cheng}}, \bibinfo {author} {\bibfnamefont
  {H.}~\bibnamefont {Ramezani}}, \bibinfo {author} {\bibfnamefont
  {Y.}~\bibnamefont {Wang}}, \ and\ \bibinfo {author} {\bibfnamefont
  {X.}~\bibnamefont {Zhang}},\ }\bibfield  {title} {\enquote {\bibinfo {title}
  {{Accessing the exceptional points of parity-time symmetric acoustics}},}\
  }\href@noop {} {\bibfield  {journal} {\bibinfo  {journal} {Nature
  communications}\ }\textbf {\bibinfo {volume} {7}},\ \bibinfo {pages} {11110}
  (\bibinfo {year} {2016})}\BibitemShut {NoStop}%
\bibitem [{\citenamefont {Xu}\ \emph {et~al.}(2020)\citenamefont {Xu},
  \citenamefont {Wu}, \citenamefont {Chen}, \citenamefont {Nassar},
  \citenamefont {Chen}, \citenamefont {Norris}, \citenamefont {Haberman},\ and\
  \citenamefont {Huang}}]{xu2020prl}%
  \BibitemOpen
  \bibfield  {author} {\bibinfo {author} {\bibfnamefont {X.}~\bibnamefont
  {Xu}}, \bibinfo {author} {\bibfnamefont {Q.}~\bibnamefont {Wu}}, \bibinfo
  {author} {\bibfnamefont {H.}~\bibnamefont {Chen}}, \bibinfo {author}
  {\bibfnamefont {H.}~\bibnamefont {Nassar}}, \bibinfo {author} {\bibfnamefont
  {Y.}~\bibnamefont {Chen}}, \bibinfo {author} {\bibfnamefont {A.}~\bibnamefont
  {Norris}}, \bibinfo {author} {\bibfnamefont {M.~R.}\ \bibnamefont
  {Haberman}}, \ and\ \bibinfo {author} {\bibfnamefont {G.}~\bibnamefont
  {Huang}},\ }\bibfield  {title} {\enquote {\bibinfo {title} {Physical
  observation of a robust acoustic pumping in waveguides with dynamic
  boundary},}\ }\href {\doibase 10.1103/PhysRevLett.125.253901} {\bibfield
  {journal} {\bibinfo  {journal} {Phys. Rev. Lett.}\ }\textbf {\bibinfo
  {volume} {125}},\ \bibinfo {pages} {253901} (\bibinfo {year}
  {2020})}\BibitemShut {NoStop}%
\bibitem [{\citenamefont {Thevamaran}\ \emph {et~al.}(2019)\citenamefont
  {Thevamaran}, \citenamefont {Branscomb}, \citenamefont {Makri}, \citenamefont
  {Anzel}, \citenamefont {Christodoulides}, \citenamefont {Kottos},\ and\
  \citenamefont {Thomas}}]{Thevamaran2019}%
  \BibitemOpen
  \bibfield  {author} {\bibinfo {author} {\bibfnamefont {R.}~\bibnamefont
  {Thevamaran}}, \bibinfo {author} {\bibfnamefont {R.~M.}\ \bibnamefont
  {Branscomb}}, \bibinfo {author} {\bibfnamefont {E.}~\bibnamefont {Makri}},
  \bibinfo {author} {\bibfnamefont {P.}~\bibnamefont {Anzel}}, \bibinfo
  {author} {\bibfnamefont {D.}~\bibnamefont {Christodoulides}}, \bibinfo
  {author} {\bibfnamefont {T.}~\bibnamefont {Kottos}}, \ and\ \bibinfo {author}
  {\bibfnamefont {E.~L.}\ \bibnamefont {Thomas}},\ }\bibfield  {title}
  {\enquote {\bibinfo {title} {Asymmetric acoustic energy transport in
  non-hermitian metamaterials},}\ }\href {\doibase 10.1121/1.5114919}
  {\bibfield  {journal} {\bibinfo  {journal} {The Journal of the Acoustical
  Society of America}\ }\textbf {\bibinfo {volume} {146}},\ \bibinfo {pages}
  {863--872} (\bibinfo {year} {2019})},\ \Eprint
  {http://arxiv.org/abs/https://doi.org/10.1121/1.5114919}
  {https://doi.org/10.1121/1.5114919} \BibitemShut {NoStop}%
\bibitem [{\citenamefont {Christensen}\ \emph {et~al.}(2016)\citenamefont
  {Christensen}, \citenamefont {Willatzen}, \citenamefont {Velasco},\ and\
  \citenamefont {Lu}}]{Christensen2016prl}%
  \BibitemOpen
  \bibfield  {author} {\bibinfo {author} {\bibfnamefont {J.}~\bibnamefont
  {Christensen}}, \bibinfo {author} {\bibfnamefont {M.}~\bibnamefont
  {Willatzen}}, \bibinfo {author} {\bibfnamefont {V.~R.}\ \bibnamefont
  {Velasco}}, \ and\ \bibinfo {author} {\bibfnamefont {M.-H.}\ \bibnamefont
  {Lu}},\ }\bibfield  {title} {\enquote {\bibinfo {title} {{Parity-Time
  Synthetic Phononic Media}},}\ }\href {\doibase
  10.1103/PhysRevLett.116.207601} {\bibfield  {journal} {\bibinfo  {journal}
  {Phys. Rev. Lett.}\ }\textbf {\bibinfo {volume} {116}},\ \bibinfo {pages}
  {207601} (\bibinfo {year} {2016})}\BibitemShut {NoStop}%
\bibitem [{\citenamefont {Hou}\ and\ \citenamefont
  {Assouar}(2018)}]{hou2018jap}%
  \BibitemOpen
  \bibfield  {author} {\bibinfo {author} {\bibfnamefont {Z.}~\bibnamefont
  {Hou}}\ and\ \bibinfo {author} {\bibfnamefont {B.}~\bibnamefont {Assouar}},\
  }\bibfield  {title} {\enquote {\bibinfo {title} {{Tunable elastic parity-time
  symmetric structure based on the shunted piezoelectric materials}},}\ }\href
  {\doibase 10.1063/1.5009129} {\bibfield  {journal} {\bibinfo  {journal}
  {Journal of Applied Physics}\ }\textbf {\bibinfo {volume} {123}},\ \bibinfo
  {pages} {85101} (\bibinfo {year} {2018})}\BibitemShut {NoStop}%
\bibitem [{\citenamefont {Merkel}, \citenamefont {Willatzen},\ and\
  \citenamefont {Christensen}(2018)}]{Merkel2018}%
  \BibitemOpen
  \bibfield  {author} {\bibinfo {author} {\bibfnamefont {A.}~\bibnamefont
  {Merkel}}, \bibinfo {author} {\bibfnamefont {M.}~\bibnamefont {Willatzen}}, \
  and\ \bibinfo {author} {\bibfnamefont {J.}~\bibnamefont {Christensen}},\
  }\bibfield  {title} {\enquote {\bibinfo {title} {{Dynamic Nonreciprocity in
  Loss-Compensated Piezophononic Media}},}\ }\href {\doibase
  10.1103/PhysRevApplied.9.034033} {\bibfield  {journal} {\bibinfo  {journal}
  {Physical Review Applied}\ }\textbf {\bibinfo {volume} {9}},\ \bibinfo
  {pages} {1--6} (\bibinfo {year} {2018})}\BibitemShut {NoStop}%
\bibitem [{\citenamefont {Hou}, \citenamefont {Ni},\ and\ \citenamefont
  {Assouar}(2018)}]{Hou2018PRApplied}%
  \BibitemOpen
  \bibfield  {author} {\bibinfo {author} {\bibfnamefont {Z.}~\bibnamefont
  {Hou}}, \bibinfo {author} {\bibfnamefont {H.}~\bibnamefont {Ni}}, \ and\
  \bibinfo {author} {\bibfnamefont {B.}~\bibnamefont {Assouar}},\ }\bibfield
  {title} {\enquote {\bibinfo {title} {Pt-symmetry for elastic negative
  refraction},}\ }\href {\doibase 10.1103/PhysRevApplied.10.044071} {\bibfield
  {journal} {\bibinfo  {journal} {Phys. Rev. Applied}\ }\textbf {\bibinfo
  {volume} {10}},\ \bibinfo {pages} {44071} (\bibinfo {year}
  {2018})}\BibitemShut {NoStop}%
\bibitem [{\citenamefont {Psiachos}\ and\ \citenamefont
  {Sigalas}(2018)}]{Psiachos2018jap}%
  \BibitemOpen
  \bibfield  {author} {\bibinfo {author} {\bibfnamefont {D.}~\bibnamefont
  {Psiachos}}\ and\ \bibinfo {author} {\bibfnamefont {M.~M.}\ \bibnamefont
  {Sigalas}},\ }\bibfield  {title} {\enquote {\bibinfo {title} {{Acoustic
  response in a one-dimensional layered pseudo-Hermitian metamaterial
  containing defects}},}\ }\href {\doibase 10.1063/1.5027457} {\bibfield
  {journal} {\bibinfo  {journal} {Journal of Applied Physics}\ }\textbf
  {\bibinfo {volume} {123}},\ \bibinfo {pages} {245109} (\bibinfo {year}
  {2018})}\BibitemShut {NoStop}%
\bibitem [{\citenamefont {Merkel}\ \emph {et~al.}(2018)\citenamefont {Merkel},
  \citenamefont {Romero-Garc\'{\i}a}, \citenamefont {Groby}, \citenamefont
  {Li},\ and\ \citenamefont {Christensen}}]{Merkel2018prb}%
  \BibitemOpen
  \bibfield  {author} {\bibinfo {author} {\bibfnamefont {A.}~\bibnamefont
  {Merkel}}, \bibinfo {author} {\bibfnamefont {V.}~\bibnamefont
  {Romero-Garc\'{\i}a}}, \bibinfo {author} {\bibfnamefont {J.-P.}\ \bibnamefont
  {Groby}}, \bibinfo {author} {\bibfnamefont {J.}~\bibnamefont {Li}}, \ and\
  \bibinfo {author} {\bibfnamefont {J.}~\bibnamefont {Christensen}},\
  }\bibfield  {title} {\enquote {\bibinfo {title} {{Unidirectional zero sonic
  reflection in passive $\mathcal{PT}$-symmetric Willis media}},}\ }\href
  {\doibase 10.1103/PhysRevB.98.201102} {\bibfield  {journal} {\bibinfo
  {journal} {Phys. Rev. B}\ }\textbf {\bibinfo {volume} {98}},\ \bibinfo
  {pages} {201102} (\bibinfo {year} {2018})}\BibitemShut {NoStop}%
\bibitem [{\citenamefont {{Chew}}(2008)}]{Chew2008}%
  \BibitemOpen
  \bibfield  {author} {\bibinfo {author} {\bibfnamefont {W.~C.}\ \bibnamefont
  {{Chew}}},\ }\bibfield  {title} {\enquote {\bibinfo {title} {A new look at
  reciprocity and energy conservation theorems in electromagnetics},}\ }\href
  {\doibase 10.1109/TAP.2008.919189} {\bibfield  {journal} {\bibinfo  {journal}
  {IEEE Transactions on Antennas and Propagation}\ }\textbf {\bibinfo {volume}
  {56}},\ \bibinfo {pages} {970--975} (\bibinfo {year} {2008})}\BibitemShut
  {NoStop}%
\bibitem [{\citenamefont {Srivastava}(2015)}]{Srivastava2015prsa}%
  \BibitemOpen
  \bibfield  {author} {\bibinfo {author} {\bibfnamefont {A.}~\bibnamefont
  {Srivastava}},\ }\bibfield  {title} {\enquote {\bibinfo {title} {{Causality
  and passivity in elastodynamics}},}\ }\href {\doibase 10.1098/rspa.2015.0256}
  {\bibfield  {journal} {\bibinfo  {journal} {Proceedings of the Royal Society
  of London A: Mathematical, Physical and Engineering Sciences}\ }\textbf
  {\bibinfo {volume} {471}} (\bibinfo {year} {2015}),\
  10.1098/rspa.2015.0256}\BibitemShut {NoStop}%
\bibitem [{\citenamefont {Pernas-Salom\'on}\ and\ \citenamefont
  {Shmuel}(2020)}]{pernassalomn2020prapplied}%
  \BibitemOpen
  \bibfield  {author} {\bibinfo {author} {\bibfnamefont {R.}~\bibnamefont
  {Pernas-Salom\'on}}\ and\ \bibinfo {author} {\bibfnamefont {G.}~\bibnamefont
  {Shmuel}},\ }\bibfield  {title} {\enquote {\bibinfo {title} {{Fundamental
  Principles for Generalized Willis Metamaterials}},}\ }\href {\doibase
  10.1103/PhysRevApplied.14.064005} {\bibfield  {journal} {\bibinfo  {journal}
  {Phys. Rev. Applied}\ }\textbf {\bibinfo {volume} {14}},\ \bibinfo {pages}
  {064005} (\bibinfo {year} {2020})}\BibitemShut {NoStop}%
\bibitem [{\citenamefont {Zhao}, \citenamefont {Longhi},\ and\ \citenamefont
  {Feng}(2015)}]{Longhi2015SR}%
  \BibitemOpen
  \bibfield  {author} {\bibinfo {author} {\bibfnamefont {H.}~\bibnamefont
  {Zhao}}, \bibinfo {author} {\bibfnamefont {S.}~\bibnamefont {Longhi}}, \ and\
  \bibinfo {author} {\bibfnamefont {L.}~\bibnamefont {Feng}},\ }\bibfield
  {title} {\enquote {\bibinfo {title} {Robust light state by quantum phase
  transition in non-hermitian optical materials},}\ }\href {\doibase
  10.1038/srep17022} {\bibfield  {journal} {\bibinfo  {journal} {Scientific
  Reports}\ }\textbf {\bibinfo {volume} {5}} (\bibinfo {year} {2015}),\
  10.1038/srep17022}\BibitemShut {NoStop}%
\bibitem [{\citenamefont {Longhi}(2017{\natexlab{b}})}]{Longhi2017jpa}%
  \BibitemOpen
  \bibfield  {author} {\bibinfo {author} {\bibfnamefont {S.}~\bibnamefont
  {Longhi}},\ }\bibfield  {title} {\enquote {\bibinfo {title} {Floquet
  exceptional points and chirality in non-hermitian hamiltonians},}\ }\href
  {\doibase 10.1088/1751-8121/aa931f} {\bibfield  {journal} {\bibinfo
  {journal} {Journal of Physics A: Mathematical and Theoretical}\ }\textbf
  {\bibinfo {volume} {50}},\ \bibinfo {pages} {505201} (\bibinfo {year}
  {2017}{\natexlab{b}})}\BibitemShut {NoStop}%
\bibitem [{\citenamefont {Lustig}\ \emph {et~al.}(2019)\citenamefont {Lustig},
  \citenamefont {Elbaz}, \citenamefont {Muhafra},\ and\ \citenamefont
  {Shmuel}}]{Lustig2019}%
  \BibitemOpen
  \bibfield  {author} {\bibinfo {author} {\bibfnamefont {B.}~\bibnamefont
  {Lustig}}, \bibinfo {author} {\bibfnamefont {G.}~\bibnamefont {Elbaz}},
  \bibinfo {author} {\bibfnamefont {A.}~\bibnamefont {Muhafra}}, \ and\
  \bibinfo {author} {\bibfnamefont {G.}~\bibnamefont {Shmuel}},\ }\bibfield
  {title} {\enquote {\bibinfo {title} {Anomalous energy transport in laminates
  with exceptional points},}\ }\href {\doibase
  https://doi.org/10.1016/j.jmps.2019.103719} {\bibfield  {journal} {\bibinfo
  {journal} {Journal of the Mechanics and Physics of Solids}\ ,\ \bibinfo
  {pages} {103719}} (\bibinfo {year} {2019})}\BibitemShut {NoStop}%
\bibitem [{\citenamefont {Mokhtari}\ \emph {et~al.}(2020)\citenamefont
  {Mokhtari}, \citenamefont {Lu}, \citenamefont {Zhou}, \citenamefont
  {Amirkhizi},\ and\ \citenamefont {Srivastava}}]{mokhtari2020scattering}%
  \BibitemOpen
  \bibfield  {author} {\bibinfo {author} {\bibfnamefont {A.~A.}\ \bibnamefont
  {Mokhtari}}, \bibinfo {author} {\bibfnamefont {Y.}~\bibnamefont {Lu}},
  \bibinfo {author} {\bibfnamefont {Q.}~\bibnamefont {Zhou}}, \bibinfo {author}
  {\bibfnamefont {A.~V.}\ \bibnamefont {Amirkhizi}}, \ and\ \bibinfo {author}
  {\bibfnamefont {A.}~\bibnamefont {Srivastava}},\ }\bibfield  {title}
  {\enquote {\bibinfo {title} {Scattering of in-plane elastic waves at
  metamaterial interfaces},}\ }\href@noop {} {\bibfield  {journal} {\bibinfo
  {journal} {International Journal of Engineering Science}\ }\textbf {\bibinfo
  {volume} {150}},\ \bibinfo {pages} {103278} (\bibinfo {year}
  {2020})}\BibitemShut {NoStop}%
\bibitem [{\citenamefont {Fishman}\ \emph {et~al.}(2024)\citenamefont
  {Fishman}, \citenamefont {Elbaz}, \citenamefont {Varma},\ and\ \citenamefont
  {Shmuel}}]{FISHMAN2024JMPS}%
  \BibitemOpen
  \bibfield  {author} {\bibinfo {author} {\bibfnamefont {A.}~\bibnamefont
  {Fishman}}, \bibinfo {author} {\bibfnamefont {G.}~\bibnamefont {Elbaz}},
  \bibinfo {author} {\bibfnamefont {T.~V.}\ \bibnamefont {Varma}}, \ and\
  \bibinfo {author} {\bibfnamefont {G.}~\bibnamefont {Shmuel}},\ }\bibfield
  {title} {\enquote {\bibinfo {title} {Third-order exceptional points and
  frozen modes in planar elastic laminates},}\ }\href {\doibase
  https://doi.org/10.1016/j.jmps.2024.105590} {\bibfield  {journal} {\bibinfo
  {journal} {Journal of the Mechanics and Physics of Solids}\ }\textbf
  {\bibinfo {volume} {186}},\ \bibinfo {pages} {105590} (\bibinfo {year}
  {2024})}\BibitemShut {NoStop}%
\bibitem [{\citenamefont {Peng}\ \emph {et~al.}(2016)\citenamefont {Peng},
  \citenamefont {{\"O}zdemir}, \citenamefont {Liertzer}, \citenamefont {Chen},
  \citenamefont {Kramer}, \citenamefont {Y{\i}lmaz}, \citenamefont {Wiersig},
  \citenamefont {Rotter},\ and\ \citenamefont {Yang}}]{Peng2016pnas}%
  \BibitemOpen
  \bibfield  {author} {\bibinfo {author} {\bibfnamefont {B.}~\bibnamefont
  {Peng}}, \bibinfo {author} {\bibfnamefont {{\c S}.~K.}\ \bibnamefont
  {{\"O}zdemir}}, \bibinfo {author} {\bibfnamefont {M.}~\bibnamefont
  {Liertzer}}, \bibinfo {author} {\bibfnamefont {W.}~\bibnamefont {Chen}},
  \bibinfo {author} {\bibfnamefont {J.}~\bibnamefont {Kramer}}, \bibinfo
  {author} {\bibfnamefont {H.}~\bibnamefont {Y{\i}lmaz}}, \bibinfo {author}
  {\bibfnamefont {J.}~\bibnamefont {Wiersig}}, \bibinfo {author} {\bibfnamefont
  {S.}~\bibnamefont {Rotter}}, \ and\ \bibinfo {author} {\bibfnamefont
  {L.}~\bibnamefont {Yang}},\ }\bibfield  {title} {\enquote {\bibinfo {title}
  {Chiral modes and directional lasing at exceptional points},}\ }\href
  {\doibase 10.1073/pnas.1603318113} {\bibfield  {journal} {\bibinfo  {journal}
  {Proceedings of the National Academy of Sciences}\ }\textbf {\bibinfo
  {volume} {113}},\ \bibinfo {pages} {6845--6850} (\bibinfo {year} {2016})},\
  \Eprint
  {http://arxiv.org/abs/https://www.pnas.org/content/113/25/6845.full.pdf}
  {https://www.pnas.org/content/113/25/6845.full.pdf} \BibitemShut {NoStop}%
\bibitem [{\citenamefont {Hodaei}\ \emph {et~al.}(2017)\citenamefont {Hodaei},
  \citenamefont {Hassan}, \citenamefont {Wittek}, \citenamefont
  {Garcia-Gracia}, \citenamefont {El-Ganainy}, \citenamefont
  {Christodoulides},\ and\ \citenamefont {Khajavikhan}}]{hodaei2017enhanced}%
  \BibitemOpen
  \bibfield  {author} {\bibinfo {author} {\bibfnamefont {H.}~\bibnamefont
  {Hodaei}}, \bibinfo {author} {\bibfnamefont {A.~U.}\ \bibnamefont {Hassan}},
  \bibinfo {author} {\bibfnamefont {S.}~\bibnamefont {Wittek}}, \bibinfo
  {author} {\bibfnamefont {H.}~\bibnamefont {Garcia-Gracia}}, \bibinfo {author}
  {\bibfnamefont {R.}~\bibnamefont {El-Ganainy}}, \bibinfo {author}
  {\bibfnamefont {D.~N.}\ \bibnamefont {Christodoulides}}, \ and\ \bibinfo
  {author} {\bibfnamefont {M.}~\bibnamefont {Khajavikhan}},\ }\bibfield
  {title} {\enquote {\bibinfo {title} {{Enhanced sensitivity at higher-order
  exceptional points}},}\ }\href@noop {} {\bibfield  {journal} {\bibinfo
  {journal} {Nature}\ }\textbf {\bibinfo {volume} {548}},\ \bibinfo {pages}
  {187} (\bibinfo {year} {2017})}\BibitemShut {NoStop}%
\bibitem [{\citenamefont {Shen}\ \emph {et~al.}(2018)\citenamefont {Shen},
  \citenamefont {Li}, \citenamefont {Peng},\ and\ \citenamefont
  {Cummer}}]{Shen2018prmat}%
  \BibitemOpen
  \bibfield  {author} {\bibinfo {author} {\bibfnamefont {C.}~\bibnamefont
  {Shen}}, \bibinfo {author} {\bibfnamefont {J.}~\bibnamefont {Li}}, \bibinfo
  {author} {\bibfnamefont {X.}~\bibnamefont {Peng}}, \ and\ \bibinfo {author}
  {\bibfnamefont {S.~A.}\ \bibnamefont {Cummer}},\ }\bibfield  {title}
  {\enquote {\bibinfo {title} {Synthetic exceptional points and unidirectional
  zero reflection in non-hermitian acoustic systems},}\ }\href {\doibase
  10.1103/PhysRevMaterials.2.125203} {\bibfield  {journal} {\bibinfo  {journal}
  {Phys. Rev. Materials}\ }\textbf {\bibinfo {volume} {2}},\ \bibinfo {pages}
  {125203} (\bibinfo {year} {2018})}\BibitemShut {NoStop}%
\bibitem [{\citenamefont {Mostafazadeh}(2002)}]{Mostafazadeh2002jmp}%
  \BibitemOpen
  \bibfield  {author} {\bibinfo {author} {\bibfnamefont {A.}~\bibnamefont
  {Mostafazadeh}},\ }\bibfield  {title} {\enquote {\bibinfo {title}
  {Pseudo-hermiticity versus pt symmetry: The necessary condition for the
  reality of the spectrum of a non-hermitian hamiltonian},}\ }\href {\doibase
  10.1063/1.1418246} {\bibfield  {journal} {\bibinfo  {journal} {Journal of
  Mathematical Physics}\ }\textbf {\bibinfo {volume} {43}},\ \bibinfo {pages}
  {205--214} (\bibinfo {year} {2002})},\ \Eprint
  {http://arxiv.org/abs/https://doi.org/10.1063/1.1418246}
  {https://doi.org/10.1063/1.1418246} \BibitemShut {NoStop}%
\bibitem [{\citenamefont {Suchkov}\ \emph {et~al.}(2016)\citenamefont
  {Suchkov}, \citenamefont {Fotsa-Ngaffo}, \citenamefont {Kenfack-Jiotsa},
  \citenamefont {Tikeng}, \citenamefont {Kofane}, \citenamefont {Kivshar},\
  and\ \citenamefont {Sukhorukov}}]{Suchkov2016njp}%
  \BibitemOpen
  \bibfield  {author} {\bibinfo {author} {\bibfnamefont {S.~V.}\ \bibnamefont
  {Suchkov}}, \bibinfo {author} {\bibfnamefont {F.}~\bibnamefont
  {Fotsa-Ngaffo}}, \bibinfo {author} {\bibfnamefont {A.}~\bibnamefont
  {Kenfack-Jiotsa}}, \bibinfo {author} {\bibfnamefont {A.~D.}\ \bibnamefont
  {Tikeng}}, \bibinfo {author} {\bibfnamefont {T.~C.}\ \bibnamefont {Kofane}},
  \bibinfo {author} {\bibfnamefont {Y.~S.}\ \bibnamefont {Kivshar}}, \ and\
  \bibinfo {author} {\bibfnamefont {A.~A.}\ \bibnamefont {Sukhorukov}},\
  }\bibfield  {title} {\enquote {\bibinfo {title} {Non-hermitian trimers:
  {PT}-symmetry versus pseudo-hermiticity},}\ }\href {\doibase
  10.1088/1367-2630/18/6/065005} {\bibfield  {journal} {\bibinfo  {journal}
  {New Journal of Physics}\ }\textbf {\bibinfo {volume} {18}},\ \bibinfo
  {pages} {065005} (\bibinfo {year} {2016})}\BibitemShut {NoStop}%
\bibitem [{\citenamefont {Griffiths}(1999)}]{griffiths1999introduction}%
  \BibitemOpen
  \bibfield  {author} {\bibinfo {author} {\bibfnamefont {D.~J.}\ \bibnamefont
  {Griffiths}},\ }\href {https://books.google.co.il/books?id=M8XvAAAAMAAJ}
  {\emph {\bibinfo {title} {{Introduction to Electrodynamics}}}}\ (\bibinfo
  {publisher} {Prentice Hall},\ \bibinfo {year} {1999})\BibitemShut {NoStop}%
\bibitem [{\citenamefont {Joannopoulos}\ \emph {et~al.}(2008)\citenamefont
  {Joannopoulos}, \citenamefont {Johnson}, \citenamefont {Winn},\ and\
  \citenamefont {Meade}}]{Joannopoulos2008book}%
  \BibitemOpen
  \bibfield  {author} {\bibinfo {author} {\bibfnamefont {J.~D.}\ \bibnamefont
  {Joannopoulos}}, \bibinfo {author} {\bibfnamefont {S.~G.}\ \bibnamefont
  {Johnson}}, \bibinfo {author} {\bibfnamefont {J.~N.}\ \bibnamefont {Winn}}, \
  and\ \bibinfo {author} {\bibfnamefont {R.~D.}\ \bibnamefont {Meade}},\ }\href
  {http://www.jstor.org/stable/j.ctvcm4gz9} {\emph {\bibinfo {title} {Photonic
  Crystals: Molding the Flow of Light - Second Edition}}},\ \bibinfo {edition}
  {rev - revised, 2}\ ed.\ (\bibinfo  {publisher} {Princeton University
  Press},\ \bibinfo {year} {2008})\BibitemShut {NoStop}%
\bibitem [{\citenamefont {Kinsler}\ \emph {et~al.}(2000)\citenamefont
  {Kinsler}, \citenamefont {Frey}, \citenamefont {Coppens},\ and\ \citenamefont
  {Sanders}}]{kinsler2000fundamentals}%
  \BibitemOpen
  \bibfield  {author} {\bibinfo {author} {\bibfnamefont {L.~E.}\ \bibnamefont
  {Kinsler}}, \bibinfo {author} {\bibfnamefont {A.~R.}\ \bibnamefont {Frey}},
  \bibinfo {author} {\bibfnamefont {A.~B.}\ \bibnamefont {Coppens}}, \ and\
  \bibinfo {author} {\bibfnamefont {J.~V.}\ \bibnamefont {Sanders}},\
  }\href@noop {} {\emph {\bibinfo {title} {Fundamentals of acoustics}}}\
  (\bibinfo  {publisher} {John Wiley \& Sons},\ \bibinfo {year}
  {2000})\BibitemShut {NoStop}%
\bibitem [{\citenamefont {Graff}(1975)}]{graff1975wave}%
  \BibitemOpen
  \bibfield  {author} {\bibinfo {author} {\bibfnamefont {K.~F.}\ \bibnamefont
  {Graff}},\ }\href {https://books.google.co.il/books?id=5cZFRwLuhdQC} {\emph
  {\bibinfo {title} {{Wave Motion in Elastic Solids}}}},\ Dover Books on
  Physics Series\ (\bibinfo  {publisher} {Dover Publications},\ \bibinfo {year}
  {1975})\BibitemShut {NoStop}%
\bibitem [{\citenamefont {Auld}(1973)}]{auld1973acoustic}%
  \BibitemOpen
  \bibfield  {author} {\bibinfo {author} {\bibfnamefont {B.}~\bibnamefont
  {Auld}},\ }\href {https://books.google.co.il/books?id=\_2MWAwAAQBAJ} {\emph
  {\bibinfo {title} {Acoustic fields and waves in solids}}},\ A
  Wiley-Interscience publication\ (\bibinfo  {publisher} {Wiley},\ \bibinfo
  {year} {1973})\BibitemShut {NoStop}%
\bibitem [{\citenamefont {Nayfeh}(1995)}]{nayfeh1995wave}%
  \BibitemOpen
  \bibfield  {author} {\bibinfo {author} {\bibfnamefont {A.}~\bibnamefont
  {Nayfeh}},\ }\href {https://books.google.co.il/books?id=8p4eAQAAIAAJ} {\emph
  {\bibinfo {title} {Wave Propagation in Layered Anisotropic Media: With
  Application to Composites}}},\ North-Holland Series in Applied Mathematics
  and Mechanics\ (\bibinfo  {publisher} {Elsevier Science},\ \bibinfo {year}
  {1995})\BibitemShut {NoStop}%
\bibitem [{\citenamefont {Long}, \citenamefont {Ren},\ and\ \citenamefont
  {Chen}(2018)}]{Long2018pnas}%
  \BibitemOpen
  \bibfield  {author} {\bibinfo {author} {\bibfnamefont {Y.}~\bibnamefont
  {Long}}, \bibinfo {author} {\bibfnamefont {J.}~\bibnamefont {Ren}}, \ and\
  \bibinfo {author} {\bibfnamefont {H.}~\bibnamefont {Chen}},\ }\bibfield
  {title} {\enquote {\bibinfo {title} {Intrinsic spin of elastic waves},}\
  }\href {\doibase 10.1073/pnas.1808534115} {\bibfield  {journal} {\bibinfo
  {journal} {Proceedings of the National Academy of Sciences}\ }\textbf
  {\bibinfo {volume} {115}},\ \bibinfo {pages} {9951--9955} (\bibinfo {year}
  {2018})},\ \Eprint
  {http://arxiv.org/abs/https://www.pnas.org/content/115/40/9951.full.pdf}
  {https://www.pnas.org/content/115/40/9951.full.pdf} \BibitemShut {NoStop}%
\bibitem [{\citenamefont {Willatzen}\ and\ \citenamefont
  {Christensen}(2014)}]{Christensen2014prb}%
  \BibitemOpen
  \bibfield  {author} {\bibinfo {author} {\bibfnamefont {M.}~\bibnamefont
  {Willatzen}}\ and\ \bibinfo {author} {\bibfnamefont {J.}~\bibnamefont
  {Christensen}},\ }\bibfield  {title} {\enquote {\bibinfo {title} {Acoustic
  gain in piezoelectric semiconductors at $\ensuremath{\varepsilon}$-near-zero
  response},}\ }\href {\doibase 10.1103/PhysRevB.89.041201} {\bibfield
  {journal} {\bibinfo  {journal} {Phys. Rev. B}\ }\textbf {\bibinfo {volume}
  {89}},\ \bibinfo {pages} {041201} (\bibinfo {year} {2014})}\BibitemShut
  {NoStop}%
\bibitem [{\citenamefont {Christensen}\ and\ \citenamefont
  {Willatzen}(2015)}]{Christensen2015acta}%
  \BibitemOpen
  \bibfield  {author} {\bibinfo {author} {\bibfnamefont {J.}~\bibnamefont
  {Christensen}}\ and\ \bibinfo {author} {\bibfnamefont {M.}~\bibnamefont
  {Willatzen}},\ }\bibfield  {title} {\enquote {\bibinfo {title} {Tunable
  broadband acoustic gain in piezoelectric semiconductors at -near-zero
  response},}\ }\href {\doibase doi:10.3813/AAA.918893} {\bibfield  {journal}
  {\bibinfo  {journal} {Acta Acustica united with Acustica}\ }\textbf {\bibinfo
  {volume} {101}},\ \bibinfo {pages} {986--992} (\bibinfo {year}
  {2015})}\BibitemShut {NoStop}%
\bibitem [{\citenamefont {Christensen}(2016)}]{Christensen2016epl}%
  \BibitemOpen
  \bibfield  {author} {\bibinfo {author} {\bibfnamefont {J.}~\bibnamefont
  {Christensen}},\ }\bibfield  {title} {\enquote {\bibinfo {title} {Coalescence
  towards exceptional contours in synthetic phononic media},}\ }\href {\doibase
  10.1209/0295-5075/114/47007} {\bibfield  {journal} {\bibinfo  {journal}
  {{EPL} (Europhysics Letters)}\ }\textbf {\bibinfo {volume} {114}},\ \bibinfo
  {pages} {47007} (\bibinfo {year} {2016})}\BibitemShut {NoStop}%
\bibitem [{\citenamefont {Gao}, \citenamefont {Willatzen},\ and\ \citenamefont
  {Christensen}(2020)}]{gao2020prl}%
  \BibitemOpen
  \bibfield  {author} {\bibinfo {author} {\bibfnamefont {P.}~\bibnamefont
  {Gao}}, \bibinfo {author} {\bibfnamefont {M.}~\bibnamefont {Willatzen}}, \
  and\ \bibinfo {author} {\bibfnamefont {J.}~\bibnamefont {Christensen}},\
  }\bibfield  {title} {\enquote {\bibinfo {title} {Anomalous topological edge
  states in non-hermitian piezophononic media},}\ }\href {\doibase
  10.1103/PhysRevLett.125.206402} {\bibfield  {journal} {\bibinfo  {journal}
  {Phys. Rev. Lett.}\ }\textbf {\bibinfo {volume} {125}},\ \bibinfo {pages}
  {206402} (\bibinfo {year} {2020})}\BibitemShut {NoStop}%
\bibitem [{\citenamefont {Gardonio}\ and\ \citenamefont
  {Casagrande}(2017)}]{GARDONIO2017}%
  \BibitemOpen
  \bibfield  {author} {\bibinfo {author} {\bibfnamefont {P.}~\bibnamefont
  {Gardonio}}\ and\ \bibinfo {author} {\bibfnamefont {D.}~\bibnamefont
  {Casagrande}},\ }\bibfield  {title} {\enquote {\bibinfo {title} {Shunted
  piezoelectric patch vibration absorber on two-dimensional thin structures:
  Tuning considerations},}\ }\href {\doibase
  https://doi.org/10.1016/j.jsv.2017.02.019} {\bibfield  {journal} {\bibinfo
  {journal} {Journal of Sound and Vibration}\ }\textbf {\bibinfo {volume}
  {395}},\ \bibinfo {pages} {26--47} (\bibinfo {year} {2017})}\BibitemShut
  {NoStop}%
\bibitem [{\citenamefont {Wu}, \citenamefont {Chen},\ and\ \citenamefont
  {Huang}(2019)}]{wu2019asymmetric}%
  \BibitemOpen
  \bibfield  {author} {\bibinfo {author} {\bibfnamefont {Q.}~\bibnamefont
  {Wu}}, \bibinfo {author} {\bibfnamefont {Y.}~\bibnamefont {Chen}}, \ and\
  \bibinfo {author} {\bibfnamefont {G.}~\bibnamefont {Huang}},\ }\bibfield
  {title} {\enquote {\bibinfo {title} {Asymmetric scattering of flexural waves
  in a parity-time symmetric metamaterial beam},}\ }\href@noop {} {\bibfield
  {journal} {\bibinfo  {journal} {The Journal of the Acoustical Society of
  America}\ }\textbf {\bibinfo {volume} {146}},\ \bibinfo {pages} {850--862}
  (\bibinfo {year} {2019})}\BibitemShut {NoStop}%
\bibitem [{\citenamefont {Marconi}\ \emph {et~al.}(2020)\citenamefont
  {Marconi}, \citenamefont {Riva}, \citenamefont {Di~Ronco}, \citenamefont
  {Cazzulani}, \citenamefont {Braghin},\ and\ \citenamefont
  {Ruzzene}}]{marconi2020prapplied}%
  \BibitemOpen
  \bibfield  {author} {\bibinfo {author} {\bibfnamefont {J.}~\bibnamefont
  {Marconi}}, \bibinfo {author} {\bibfnamefont {E.}~\bibnamefont {Riva}},
  \bibinfo {author} {\bibfnamefont {M.}~\bibnamefont {Di~Ronco}}, \bibinfo
  {author} {\bibfnamefont {G.}~\bibnamefont {Cazzulani}}, \bibinfo {author}
  {\bibfnamefont {F.}~\bibnamefont {Braghin}}, \ and\ \bibinfo {author}
  {\bibfnamefont {M.}~\bibnamefont {Ruzzene}},\ }\bibfield  {title} {\enquote
  {\bibinfo {title} {Experimental observation of nonreciprocal band gaps in a
  space-time-modulated beam using a shunted piezoelectric array},}\ }\href
  {\doibase 10.1103/PhysRevApplied.13.031001} {\bibfield  {journal} {\bibinfo
  {journal} {Phys. Rev. Applied}\ }\textbf {\bibinfo {volume} {13}},\ \bibinfo
  {pages} {031001} (\bibinfo {year} {2020})}\BibitemShut {NoStop}%
\bibitem [{\citenamefont {Xia}\ \emph {et~al.}(2021)\citenamefont {Xia},
  \citenamefont {Riva}, \citenamefont {Rosa}, \citenamefont {Cazzulani},
  \citenamefont {Erturk}, \citenamefont {Braghin},\ and\ \citenamefont
  {Ruzzene}}]{Xia2021prl}%
  \BibitemOpen
  \bibfield  {author} {\bibinfo {author} {\bibfnamefont {Y.}~\bibnamefont
  {Xia}}, \bibinfo {author} {\bibfnamefont {E.}~\bibnamefont {Riva}}, \bibinfo
  {author} {\bibfnamefont {M.~I.~N.}\ \bibnamefont {Rosa}}, \bibinfo {author}
  {\bibfnamefont {G.}~\bibnamefont {Cazzulani}}, \bibinfo {author}
  {\bibfnamefont {A.}~\bibnamefont {Erturk}}, \bibinfo {author} {\bibfnamefont
  {F.}~\bibnamefont {Braghin}}, \ and\ \bibinfo {author} {\bibfnamefont
  {M.}~\bibnamefont {Ruzzene}},\ }\bibfield  {title} {\enquote {\bibinfo
  {title} {Experimental observation of temporal pumping in electromechanical
  waveguides},}\ }\href {\doibase 10.1103/PhysRevLett.126.095501} {\bibfield
  {journal} {\bibinfo  {journal} {Phys. Rev. Lett.}\ }\textbf {\bibinfo
  {volume} {126}},\ \bibinfo {pages} {095501} (\bibinfo {year}
  {2021})}\BibitemShut {NoStop}%
\bibitem [{\citenamefont {Thomes}, \citenamefont {Rosa},\ and\ \citenamefont
  {Erturk}(2024)}]{Thomes2024apl}%
  \BibitemOpen
  \bibfield  {author} {\bibinfo {author} {\bibfnamefont {R.~L.}\ \bibnamefont
  {Thomes}}, \bibinfo {author} {\bibfnamefont {M.~I.~N.}\ \bibnamefont {Rosa}},
  \ and\ \bibinfo {author} {\bibfnamefont {A.}~\bibnamefont {Erturk}},\
  }\bibfield  {title} {\enquote {\bibinfo {title} {{Experimental realization of
  tunable exceptional points in a resonant non-Hermitian piezoelectrically
  coupled waveguide}},}\ }\href {\doibase 10.1063/5.0183401} {\bibfield
  {journal} {\bibinfo  {journal} {Applied Physics Letters}\ }\textbf {\bibinfo
  {volume} {124}},\ \bibinfo {pages} {061702} (\bibinfo {year} {2024})},\
  \Eprint
  {http://arxiv.org/abs/https://pubs.aip.org/aip/apl/article-pdf/doi/10.1063/5.0183401/19616502/061702\_1\_5.0183401.pdf}
  {https://pubs.aip.org/aip/apl/article-pdf/doi/10.1063/5.0183401/19616502/061702\_1\_5.0183401.pdf}
  \BibitemShut {NoStop}%
\bibitem [{\citenamefont {Nassar}\ \emph {et~al.}(2017)\citenamefont {Nassar},
  \citenamefont {Xu}, \citenamefont {Norris},\ and\ \citenamefont
  {Huang}}]{nassar2017modulated}%
  \BibitemOpen
  \bibfield  {author} {\bibinfo {author} {\bibfnamefont {H.}~\bibnamefont
  {Nassar}}, \bibinfo {author} {\bibfnamefont {X.~C.}\ \bibnamefont {Xu}},
  \bibinfo {author} {\bibfnamefont {A.~N.}\ \bibnamefont {Norris}}, \ and\
  \bibinfo {author} {\bibfnamefont {G.~L.}\ \bibnamefont {Huang}},\ }\bibfield
  {title} {\enquote {\bibinfo {title} {{Modulated phononic crystals:
  Non-reciprocal wave propagation and Willis materials}},}\ }\href@noop {}
  {\bibfield  {journal} {\bibinfo  {journal} {Journal of the Mechanics and
  Physics of Solids}\ }\textbf {\bibinfo {volume} {101}},\ \bibinfo {pages}
  {10--29} (\bibinfo {year} {2017})}\BibitemShut {NoStop}%
\bibitem [{\citenamefont {Torrent}, \citenamefont {Parnell},\ and\
  \citenamefont {Norris}(2018)}]{torrent2018loss}%
  \BibitemOpen
  \bibfield  {author} {\bibinfo {author} {\bibfnamefont {D.}~\bibnamefont
  {Torrent}}, \bibinfo {author} {\bibfnamefont {W.~J.}\ \bibnamefont
  {Parnell}}, \ and\ \bibinfo {author} {\bibfnamefont {A.~N.}\ \bibnamefont
  {Norris}},\ }\bibfield  {title} {\enquote {\bibinfo {title} {Loss
  compensation in time-dependent elastic metamaterials},}\ }\href@noop {}
  {\bibfield  {journal} {\bibinfo  {journal} {Physical Review B}\ }\textbf
  {\bibinfo {volume} {97}},\ \bibinfo {pages} {014105} (\bibinfo {year}
  {2018})}\BibitemShut {NoStop}%
\bibitem [{\citenamefont {Nassar}\ \emph {et~al.}(2020)\citenamefont {Nassar},
  \citenamefont {Yousefzadeh}, \citenamefont {Fleury}, \citenamefont {Ruzzene},
  \citenamefont {Al{\`u}}, \citenamefont {Daraio}, \citenamefont {Norris},
  \citenamefont {Huang},\ and\ \citenamefont
  {Haberman}}]{nassar2020nonreciprocity}%
  \BibitemOpen
  \bibfield  {author} {\bibinfo {author} {\bibfnamefont {H.}~\bibnamefont
  {Nassar}}, \bibinfo {author} {\bibfnamefont {B.}~\bibnamefont {Yousefzadeh}},
  \bibinfo {author} {\bibfnamefont {R.}~\bibnamefont {Fleury}}, \bibinfo
  {author} {\bibfnamefont {M.}~\bibnamefont {Ruzzene}}, \bibinfo {author}
  {\bibfnamefont {A.}~\bibnamefont {Al{\`u}}}, \bibinfo {author} {\bibfnamefont
  {C.}~\bibnamefont {Daraio}}, \bibinfo {author} {\bibfnamefont {A.~N.}\
  \bibnamefont {Norris}}, \bibinfo {author} {\bibfnamefont {G.}~\bibnamefont
  {Huang}}, \ and\ \bibinfo {author} {\bibfnamefont {M.~R.}\ \bibnamefont
  {Haberman}},\ }\bibfield  {title} {\enquote {\bibinfo {title} {Nonreciprocity
  in acoustic and elastic materials},}\ }\href@noop {} {\bibfield  {journal}
  {\bibinfo  {journal} {Nature Reviews Materials}\ }\textbf {\bibinfo {volume}
  {5}},\ \bibinfo {pages} {667--685} (\bibinfo {year} {2020})}\BibitemShut
  {NoStop}%
\bibitem [{\citenamefont {Willis}(1981)}]{willis1981avariational}%
  \BibitemOpen
  \bibfield  {author} {\bibinfo {author} {\bibfnamefont {J.~R.}\ \bibnamefont
  {Willis}},\ }\bibfield  {title} {\enquote {\bibinfo {title} {{Variational and
  related methods for the overall properties of composites}},}\ }\href@noop {}
  {\bibfield  {journal} {\bibinfo  {journal} {Advances in applied mechanics}\
  }\textbf {\bibinfo {volume} {21}},\ \bibinfo {pages} {1--78} (\bibinfo {year}
  {1981})}\BibitemShut {NoStop}%
\bibitem [{\citenamefont {Willis}(2011)}]{willis2011effective}%
  \BibitemOpen
  \bibfield  {author} {\bibinfo {author} {\bibfnamefont {J.~R.}\ \bibnamefont
  {Willis}},\ }\bibfield  {title} {\enquote {\bibinfo {title} {{Effective
  constitutive relations for waves in composites and metamaterials}},}\
  }\href@noop {} {\bibfield  {journal} {\bibinfo  {journal} {Proceedings of the
  Royal Society A: Mathematical, Physical and Engineering Science}\ }\textbf
  {\bibinfo {volume} {467}},\ \bibinfo {pages} {1865--1879} (\bibinfo {year}
  {2011})}\BibitemShut {NoStop}%
\bibitem [{\citenamefont {Willis}(2012)}]{WILLIS2012MRC}%
  \BibitemOpen
  \bibfield  {author} {\bibinfo {author} {\bibfnamefont {J.~R.}\ \bibnamefont
  {Willis}},\ }\bibfield  {title} {\enquote {\bibinfo {title} {The construction
  of effective relations for waves in a composite},}\ }\href {\doibase
  https://doi.org/10.1016/j.crme.2012.02.001} {\bibfield  {journal} {\bibinfo
  {journal} {Comptes Rendus M{\'e}canique}\ }\textbf {\bibinfo {volume}
  {340}},\ \bibinfo {pages} {181 -- 192} (\bibinfo {year} {2012})},\ \bibinfo
  {note} {recent Advances in Micromechanics of Materials}\BibitemShut {NoStop}%
\bibitem [{Note1()}]{Note1}%
  \BibitemOpen
  \bibinfo {note} {Willis uses the term self-adjoint\cite {
  willis2011effective,WILLIS2012MRC} rather than Hermitian.}\BibitemShut
  {Stop}%
\bibitem [{\citenamefont {Shmuel}\ and\ \citenamefont
  {Moiseyev}(2020)}]{shmuel2020prapplied}%
  \BibitemOpen
  \bibfield  {author} {\bibinfo {author} {\bibfnamefont {G.}~\bibnamefont
  {Shmuel}}\ and\ \bibinfo {author} {\bibfnamefont {N.}~\bibnamefont
  {Moiseyev}},\ }\bibfield  {title} {\enquote {\bibinfo {title} {Linking scalar
  elastodynamics and non-hermitian quantum mechanics},}\ }\href {\doibase
  10.1103/PhysRevApplied.13.024074} {\bibfield  {journal} {\bibinfo  {journal}
  {Phys. Rev. Applied}\ }\textbf {\bibinfo {volume} {13}},\ \bibinfo {pages}
  {024074} (\bibinfo {year} {2020})}\BibitemShut {NoStop}%
\bibitem [{\citenamefont {Dom\'{\i}nguez-Rocha}\ \emph
  {et~al.}(2020)\citenamefont {Dom\'{\i}nguez-Rocha}, \citenamefont
  {Thevamaran}, \citenamefont {Ellis},\ and\ \citenamefont
  {Kottos}}]{rocha2020PRApplied}%
  \BibitemOpen
  \bibfield  {author} {\bibinfo {author} {\bibfnamefont {V.}~\bibnamefont
  {Dom\'{\i}nguez-Rocha}}, \bibinfo {author} {\bibfnamefont {R.}~\bibnamefont
  {Thevamaran}}, \bibinfo {author} {\bibfnamefont {F.}~\bibnamefont {Ellis}}, \
  and\ \bibinfo {author} {\bibfnamefont {T.}~\bibnamefont {Kottos}},\
  }\bibfield  {title} {\enquote {\bibinfo {title} {Environmentally induced
  exceptional points in elastodynamics},}\ }\href {\doibase
  10.1103/PhysRevApplied.13.014060} {\bibfield  {journal} {\bibinfo  {journal}
  {Phys. Rev. Appl.}\ }\textbf {\bibinfo {volume} {13}},\ \bibinfo {pages}
  {014060} (\bibinfo {year} {2020})}\BibitemShut {NoStop}%
\bibitem [{\citenamefont {Ogden}(1997)}]{ogden97book}%
  \BibitemOpen
  \bibfield  {author} {\bibinfo {author} {\bibfnamefont {R.~W.}\ \bibnamefont
  {Ogden}},\ }\href@noop {} {\emph {\bibinfo {title} {{Non-Linear Elastic
  Deformations}}}}\ (\bibinfo  {publisher} {Dover Publications},\ \bibinfo
  {address} {New York},\ \bibinfo {year} {1997})\BibitemShut {NoStop}%
\bibitem [{\citenamefont {Scheibner}\ \emph {et~al.}(2020)\citenamefont
  {Scheibner}, \citenamefont {Souslov}, \citenamefont {Banerjee}, \citenamefont
  {Surowka}, \citenamefont {Irvine},\ and\ \citenamefont
  {Vitelli}}]{scheibner2020odd}%
  \BibitemOpen
  \bibfield  {author} {\bibinfo {author} {\bibfnamefont {C.}~\bibnamefont
  {Scheibner}}, \bibinfo {author} {\bibfnamefont {A.}~\bibnamefont {Souslov}},
  \bibinfo {author} {\bibfnamefont {D.}~\bibnamefont {Banerjee}}, \bibinfo
  {author} {\bibfnamefont {P.}~\bibnamefont {Surowka}}, \bibinfo {author}
  {\bibfnamefont {W.~T.}\ \bibnamefont {Irvine}}, \ and\ \bibinfo {author}
  {\bibfnamefont {V.}~\bibnamefont {Vitelli}},\ }\bibfield  {title} {\enquote
  {\bibinfo {title} {Odd elasticity},}\ }\href@noop {} {\bibfield  {journal}
  {\bibinfo  {journal} {Nature Physics}\ }\textbf {\bibinfo {volume} {16}},\
  \bibinfo {pages} {475--480} (\bibinfo {year} {2020})}\BibitemShut {NoStop}%
\bibitem [{\citenamefont {Scheibner}, \citenamefont {Irvine},\ and\
  \citenamefont {Vitelli}(2020)}]{Scheibner2020prl}%
  \BibitemOpen
  \bibfield  {author} {\bibinfo {author} {\bibfnamefont {C.}~\bibnamefont
  {Scheibner}}, \bibinfo {author} {\bibfnamefont {W.~T.~M.}\ \bibnamefont
  {Irvine}}, \ and\ \bibinfo {author} {\bibfnamefont {V.}~\bibnamefont
  {Vitelli}},\ }\bibfield  {title} {\enquote {\bibinfo {title} {Non-hermitian
  band topology and skin modes in active elastic media},}\ }\href {\doibase
  10.1103/PhysRevLett.125.118001} {\bibfield  {journal} {\bibinfo  {journal}
  {Phys. Rev. Lett.}\ }\textbf {\bibinfo {volume} {125}},\ \bibinfo {pages}
  {118001} (\bibinfo {year} {2020})}\BibitemShut {NoStop}%
\bibitem [{\citenamefont {Brandenbourger}\ \emph {et~al.}(2019)\citenamefont
  {Brandenbourger}, \citenamefont {Locsin}, \citenamefont {Lerner},\ and\
  \citenamefont {Coulais}}]{brandenbourger2019non}%
  \BibitemOpen
  \bibfield  {author} {\bibinfo {author} {\bibfnamefont {M.}~\bibnamefont
  {Brandenbourger}}, \bibinfo {author} {\bibfnamefont {X.}~\bibnamefont
  {Locsin}}, \bibinfo {author} {\bibfnamefont {E.}~\bibnamefont {Lerner}}, \
  and\ \bibinfo {author} {\bibfnamefont {C.}~\bibnamefont {Coulais}},\
  }\bibfield  {title} {\enquote {\bibinfo {title} {Non-reciprocal robotic
  metamaterials},}\ }\href@noop {} {\bibfield  {journal} {\bibinfo  {journal}
  {Nature communications}\ }\textbf {\bibinfo {volume} {10}},\ \bibinfo {pages}
  {4608} (\bibinfo {year} {2019})}\BibitemShut {NoStop}%
\bibitem [{\citenamefont {Chen}\ \emph {et~al.}(2021)\citenamefont {Chen},
  \citenamefont {Li}, \citenamefont {Scheibner}, \citenamefont {Vitelli},\ and\
  \citenamefont {Huang}}]{chen2021realization}%
  \BibitemOpen
  \bibfield  {author} {\bibinfo {author} {\bibfnamefont {Y.}~\bibnamefont
  {Chen}}, \bibinfo {author} {\bibfnamefont {X.}~\bibnamefont {Li}}, \bibinfo
  {author} {\bibfnamefont {C.}~\bibnamefont {Scheibner}}, \bibinfo {author}
  {\bibfnamefont {V.}~\bibnamefont {Vitelli}}, \ and\ \bibinfo {author}
  {\bibfnamefont {G.}~\bibnamefont {Huang}},\ }\bibfield  {title} {\enquote
  {\bibinfo {title} {Realization of active metamaterials with odd micropolar
  elasticity},}\ }\href@noop {} {\bibfield  {journal} {\bibinfo  {journal}
  {Nature communications}\ }\textbf {\bibinfo {volume} {12}},\ \bibinfo {pages}
  {5935} (\bibinfo {year} {2021})}\BibitemShut {NoStop}%
\bibitem [{\citenamefont {Gao}, \citenamefont {Qu},\ and\ \citenamefont
  {Christensen}(2022)}]{gao2022non}%
  \BibitemOpen
  \bibfield  {author} {\bibinfo {author} {\bibfnamefont {P.}~\bibnamefont
  {Gao}}, \bibinfo {author} {\bibfnamefont {Y.}~\bibnamefont {Qu}}, \ and\
  \bibinfo {author} {\bibfnamefont {J.}~\bibnamefont {Christensen}},\
  }\bibfield  {title} {\enquote {\bibinfo {title} {Non-hermitian elastodynamics
  in gyro-odd continuum media},}\ }\href@noop {} {\bibfield  {journal}
  {\bibinfo  {journal} {Communications Materials}\ }\textbf {\bibinfo {volume}
  {3}},\ \bibinfo {pages} {74} (\bibinfo {year} {2022})}\BibitemShut {NoStop}%
\bibitem [{\citenamefont {Wang}, \citenamefont {Meng},\ and\ \citenamefont
  {Chen}(2023)}]{wang2023non}%
  \BibitemOpen
  \bibfield  {author} {\bibinfo {author} {\bibfnamefont {A.}~\bibnamefont
  {Wang}}, \bibinfo {author} {\bibfnamefont {Z.}~\bibnamefont {Meng}}, \ and\
  \bibinfo {author} {\bibfnamefont {C.~Q.}\ \bibnamefont {Chen}},\ }\bibfield
  {title} {\enquote {\bibinfo {title} {Non-hermitian topology in static
  mechanical metamaterials},}\ }\href@noop {} {\bibfield  {journal} {\bibinfo
  {journal} {Science advances}\ }\textbf {\bibinfo {volume} {9}},\ \bibinfo
  {pages} {eadf7299} (\bibinfo {year} {2023})}\BibitemShut {NoStop}%
\bibitem [{\citenamefont {Wu}\ \emph {et~al.}(2023)\citenamefont {Wu},
  \citenamefont {Xu}, \citenamefont {Qian}, \citenamefont {Wang}, \citenamefont
  {Zhu}, \citenamefont {Yan}, \citenamefont {Ma}, \citenamefont {Chen},\ and\
  \citenamefont {Huang}}]{wu2023active}%
  \BibitemOpen
  \bibfield  {author} {\bibinfo {author} {\bibfnamefont {Q.}~\bibnamefont
  {Wu}}, \bibinfo {author} {\bibfnamefont {X.}~\bibnamefont {Xu}}, \bibinfo
  {author} {\bibfnamefont {H.}~\bibnamefont {Qian}}, \bibinfo {author}
  {\bibfnamefont {S.}~\bibnamefont {Wang}}, \bibinfo {author} {\bibfnamefont
  {R.}~\bibnamefont {Zhu}}, \bibinfo {author} {\bibfnamefont {Z.}~\bibnamefont
  {Yan}}, \bibinfo {author} {\bibfnamefont {H.}~\bibnamefont {Ma}}, \bibinfo
  {author} {\bibfnamefont {Y.}~\bibnamefont {Chen}}, \ and\ \bibinfo {author}
  {\bibfnamefont {G.}~\bibnamefont {Huang}},\ }\bibfield  {title} {\enquote
  {\bibinfo {title} {Active metamaterials for realizing odd mass density},}\
  }\href@noop {} {\bibfield  {journal} {\bibinfo  {journal} {Proceedings of the
  National Academy of Sciences}\ }\textbf {\bibinfo {volume} {120}},\ \bibinfo
  {pages} {e2209829120} (\bibinfo {year} {2023})}\BibitemShut {NoStop}%
\bibitem [{\citenamefont {Willis}(1997)}]{willis1997dynamics}%
  \BibitemOpen
  \bibfield  {author} {\bibinfo {author} {\bibfnamefont {J.~R.}\ \bibnamefont
  {Willis}},\ }\bibfield  {title} {\enquote {\bibinfo {title} {{Dynamics of
  composites}},}\ }in\ \href@noop {} {\emph {\bibinfo {booktitle} {Continuum
  micromechanics}}}\ (\bibinfo {organization} {Springer-Verlag New York,
  Inc.},\ \bibinfo {year} {1997})\ pp.\ \bibinfo {pages} {265--290}\BibitemShut
  {NoStop}%
\bibitem [{\citenamefont {Sieck}, \citenamefont {Al{\`{u}}},\ and\
  \citenamefont {Haberman}(2017)}]{Sieck2017prb}%
  \BibitemOpen
  \bibfield  {author} {\bibinfo {author} {\bibfnamefont {C.~F.}\ \bibnamefont
  {Sieck}}, \bibinfo {author} {\bibfnamefont {A.}~\bibnamefont {Al{\`{u}}}}, \
  and\ \bibinfo {author} {\bibfnamefont {M.~R.}\ \bibnamefont {Haberman}},\
  }\bibfield  {title} {\enquote {\bibinfo {title} {{Origins of Willis coupling
  and acoustic bianisotropy in acoustic metamaterials through source-driven
  homogenization}},}\ }\href {\doibase 10.1103/PhysRevB.96.104303} {\bibfield
  {journal} {\bibinfo  {journal} {Phys. Rev. B}\ }\textbf {\bibinfo {volume}
  {96}},\ \bibinfo {pages} {104303} (\bibinfo {year} {2017})}\BibitemShut
  {NoStop}%
\bibitem [{\citenamefont {Milton}, \citenamefont {Briane},\ and\ \citenamefont
  {Willis}(2006)}]{milton06cloaking}%
  \BibitemOpen
  \bibfield  {author} {\bibinfo {author} {\bibfnamefont {G.~W.}\ \bibnamefont
  {Milton}}, \bibinfo {author} {\bibfnamefont {M.}~\bibnamefont {Briane}}, \
  and\ \bibinfo {author} {\bibfnamefont {J.~R.}\ \bibnamefont {Willis}},\
  }\bibfield  {title} {\enquote {\bibinfo {title} {{On cloaking for elasticity
  and physical equations with a transformation invariant form}},}\ }\href
  {http://stacks.iop.org/1367-2630/8/i=10/a=248} {\bibfield  {journal}
  {\bibinfo  {journal} {New J. Phys.}\ }\textbf {\bibinfo {volume} {8}},\
  \bibinfo {pages} {248} (\bibinfo {year} {2006})}\BibitemShut {NoStop}%
\bibitem [{\citenamefont {Milton}(2020)}]{Milton2020II}%
  \BibitemOpen
  \bibfield  {author} {\bibinfo {author} {\bibfnamefont {G.~W.}\ \bibnamefont
  {Milton}},\ }\bibfield  {title} {\enquote {\bibinfo {title} {A unifying
  perspective on linear continuum equations prevalent in physics. part ii:
  Canonical forms for time-harmonic equations},}\ }\href@noop {} {\bibfield
  {journal} {\bibinfo  {journal} {arXiv: Analysis of PDEs}\ } (\bibinfo {year}
  {2020})}\BibitemShut {NoStop}%
\bibitem [{\citenamefont {Chen}\ \emph {et~al.}(2020)\citenamefont {Chen},
  \citenamefont {Li}, \citenamefont {Hu}, \citenamefont {Haberman},\ and\
  \citenamefont {Huang}}]{Chen2020nc}%
  \BibitemOpen
  \bibfield  {author} {\bibinfo {author} {\bibfnamefont {Y.}~\bibnamefont
  {Chen}}, \bibinfo {author} {\bibfnamefont {X.}~\bibnamefont {Li}}, \bibinfo
  {author} {\bibfnamefont {G.}~\bibnamefont {Hu}}, \bibinfo {author}
  {\bibfnamefont {M.~R.}\ \bibnamefont {Haberman}}, \ and\ \bibinfo {author}
  {\bibfnamefont {G.}~\bibnamefont {Huang}},\ }\bibfield  {title} {\enquote
  {\bibinfo {title} {{An active mechanical Willis meta-layer with asymmetric
  polarizabilities}},}\ }\href {\doibase 10.1038/s41467-020-17529-2} {\bibfield
   {journal} {\bibinfo  {journal} {Nature Communications}\ }\textbf {\bibinfo
  {volume} {11}},\ \bibinfo {pages} {3681} (\bibinfo {year}
  {2020})}\BibitemShut {NoStop}%
\bibitem [{\citenamefont {Liu}\ \emph {et~al.}(2019)\citenamefont {Liu},
  \citenamefont {Liang}, \citenamefont {Zhu}, \citenamefont {Xia},
  \citenamefont {Mondain-Monval}, \citenamefont {Brunet}, \citenamefont
  {Al\`u},\ and\ \citenamefont {Li}}]{Liu2019prx}%
  \BibitemOpen
  \bibfield  {author} {\bibinfo {author} {\bibfnamefont {Y.}~\bibnamefont
  {Liu}}, \bibinfo {author} {\bibfnamefont {Z.}~\bibnamefont {Liang}}, \bibinfo
  {author} {\bibfnamefont {J.}~\bibnamefont {Zhu}}, \bibinfo {author}
  {\bibfnamefont {L.}~\bibnamefont {Xia}}, \bibinfo {author} {\bibfnamefont
  {O.}~\bibnamefont {Mondain-Monval}}, \bibinfo {author} {\bibfnamefont
  {T.}~\bibnamefont {Brunet}}, \bibinfo {author} {\bibfnamefont
  {A.}~\bibnamefont {Al\`u}}, \ and\ \bibinfo {author} {\bibfnamefont
  {J.}~\bibnamefont {Li}},\ }\bibfield  {title} {\enquote {\bibinfo {title}
  {Willis metamaterial on a structured beam},}\ }\href {\doibase
  10.1103/PhysRevX.9.011040} {\bibfield  {journal} {\bibinfo  {journal} {Phys.
  Rev. X}\ }\textbf {\bibinfo {volume} {9}},\ \bibinfo {pages} {011040}
  (\bibinfo {year} {2019})}\BibitemShut {NoStop}%
\bibitem [{\citenamefont {Psiachos}\ and\ \citenamefont
  {Sigalas}(2019)}]{Psiachos2019prb}%
  \BibitemOpen
  \bibfield  {author} {\bibinfo {author} {\bibfnamefont {D.}~\bibnamefont
  {Psiachos}}\ and\ \bibinfo {author} {\bibfnamefont {M.~M.}\ \bibnamefont
  {Sigalas}},\ }\bibfield  {title} {\enquote {\bibinfo {title} {Tailoring
  one-dimensional layered metamaterials to achieve unidirectional transmission
  and reflection},}\ }\href@noop {} {\bibfield  {journal} {\bibinfo  {journal}
  {{PHYSICAL REVIEW B}}\ }\textbf {\bibinfo {volume} {{99}}} (\bibinfo {year}
  {{2019}})}\BibitemShut {NoStop}%
\bibitem [{\citenamefont {Joseph}\ and\ \citenamefont
  {Craster}(2015)}]{joseph2015WM}%
  \BibitemOpen
  \bibfield  {author} {\bibinfo {author} {\bibfnamefont {L.~M.}\ \bibnamefont
  {Joseph}}\ and\ \bibinfo {author} {\bibfnamefont {R.~V.}\ \bibnamefont
  {Craster}},\ }\bibfield  {title} {\enquote {\bibinfo {title} {{Reflection
  from a semi-infinite stack of layers using homogenization}},}\ }\href
  {\doibase https://doi.org/10.1016/j.wavemoti.2014.12.003} {\bibfield
  {journal} {\bibinfo  {journal} {Wave Motion}\ }\textbf {\bibinfo {volume}
  {54}},\ \bibinfo {pages} {145--156} (\bibinfo {year} {2015})}\BibitemShut
  {NoStop}%
\bibitem [{\citenamefont {Figotin}\ and\ \citenamefont
  {Vitebskiy}(2003)}]{Figotin2003pre}%
  \BibitemOpen
  \bibfield  {author} {\bibinfo {author} {\bibfnamefont {A.}~\bibnamefont
  {Figotin}}\ and\ \bibinfo {author} {\bibfnamefont {I.}~\bibnamefont
  {Vitebskiy}},\ }\bibfield  {title} {\enquote {\bibinfo {title} {Oblique
  frozen modes in periodic layered media},}\ }\href {\doibase
  10.1103/PhysRevE.68.036609} {\bibfield  {journal} {\bibinfo  {journal} {Phys.
  Rev. E}\ }\textbf {\bibinfo {volume} {68}},\ \bibinfo {pages} {036609}
  (\bibinfo {year} {2003})}\BibitemShut {NoStop}%
\bibitem [{\citenamefont {Heiss}(2016)}]{Heiss2016yt}%
  \BibitemOpen
  \bibfield  {author} {\bibinfo {author} {\bibfnamefont {D.}~\bibnamefont
  {Heiss}},\ }\bibfield  {title} {\enquote {\bibinfo {title} {Circling
  exceptional points},}\ }\href {\doibase 10.1038/nphys3864} {\bibfield
  {journal} {\bibinfo  {journal} {Nature Physics}\ }\textbf {\bibinfo {volume}
  {12}},\ \bibinfo {pages} {823--824} (\bibinfo {year} {2016})}\BibitemShut
  {NoStop}%
\bibitem [{\citenamefont {Doppler}\ \emph {et~al.}(2016)\citenamefont
  {Doppler}, \citenamefont {Mailybaev}, \citenamefont {B{\"{o}}hm},
  \citenamefont {Kuhl}, \citenamefont {Girschik}, \citenamefont {Libisch},
  \citenamefont {Milburn}, \citenamefont {Rabl}, \citenamefont {Moiseyev},\
  and\ \citenamefont {Rotter}}]{Doppler2016nature}%
  \BibitemOpen
  \bibfield  {author} {\bibinfo {author} {\bibfnamefont {J.}~\bibnamefont
  {Doppler}}, \bibinfo {author} {\bibfnamefont {A.~A.}\ \bibnamefont
  {Mailybaev}}, \bibinfo {author} {\bibfnamefont {J.}~\bibnamefont
  {B{\"{o}}hm}}, \bibinfo {author} {\bibfnamefont {U.}~\bibnamefont {Kuhl}},
  \bibinfo {author} {\bibfnamefont {A.}~\bibnamefont {Girschik}}, \bibinfo
  {author} {\bibfnamefont {F.}~\bibnamefont {Libisch}}, \bibinfo {author}
  {\bibfnamefont {T.~J.}\ \bibnamefont {Milburn}}, \bibinfo {author}
  {\bibfnamefont {P.}~\bibnamefont {Rabl}}, \bibinfo {author} {\bibfnamefont
  {N.}~\bibnamefont {Moiseyev}}, \ and\ \bibinfo {author} {\bibfnamefont
  {S.}~\bibnamefont {Rotter}},\ }\bibfield  {title} {\enquote {\bibinfo {title}
  {{Dynamically encircling an exceptional point for asymmetric mode
  switching}},}\ }\href {https://doi.org/10.1038/nature18605} {\bibfield
  {journal} {\bibinfo  {journal} {Nature}\ }\textbf {\bibinfo {volume} {537}},\
  \bibinfo {pages} {76 EP --} (\bibinfo {year} {2016})}\BibitemShut {NoStop}%
\bibitem [{\citenamefont {Xu}\ \emph {et~al.}(2016)\citenamefont {Xu},
  \citenamefont {Mason}, \citenamefont {Jiang},\ and\ \citenamefont
  {Harris}}]{Xu2016nature}%
  \BibitemOpen
  \bibfield  {author} {\bibinfo {author} {\bibfnamefont {H.}~\bibnamefont
  {Xu}}, \bibinfo {author} {\bibfnamefont {D.}~\bibnamefont {Mason}}, \bibinfo
  {author} {\bibfnamefont {L.}~\bibnamefont {Jiang}}, \ and\ \bibinfo {author}
  {\bibfnamefont {J.~G.~E.}\ \bibnamefont {Harris}},\ }\bibfield  {title}
  {\enquote {\bibinfo {title} {Topological energy transfer in an optomechanical
  system with exceptional points},}\ }\href
  {https://doi.org/10.1038/nature18604} {\bibfield  {journal} {\bibinfo
  {journal} {Nature}\ }\textbf {\bibinfo {volume} {537}},\ \bibinfo {pages} {80
  EP --} (\bibinfo {year} {2016})}\BibitemShut {NoStop}%
\bibitem [{\citenamefont {Elbaz}\ \emph {et~al.}(2022)\citenamefont {Elbaz},
  \citenamefont {Pick}, \citenamefont {Moiseyev},\ and\ \citenamefont
  {Shmuel}}]{elbaz2022jphysd}%
  \BibitemOpen
  \bibfield  {author} {\bibinfo {author} {\bibfnamefont {G.}~\bibnamefont
  {Elbaz}}, \bibinfo {author} {\bibfnamefont {A.}~\bibnamefont {Pick}},
  \bibinfo {author} {\bibfnamefont {N.}~\bibnamefont {Moiseyev}}, \ and\
  \bibinfo {author} {\bibfnamefont {G.}~\bibnamefont {Shmuel}},\ }\bibfield
  {title} {\enquote {\bibinfo {title} {Encircling exceptional points of bloch
  waves: mode conversion and anomalous scattering},}\ }\href
  {http://iopscience.iop.org/article/10.1088/1361-6463/ac5859} {\bibfield
  {journal} {\bibinfo  {journal} {Journal of Physics D: Applied Physics}\ }
  (\bibinfo {year} {2022})}\BibitemShut {NoStop}%
\bibitem [{\citenamefont {Lu}\ and\ \citenamefont
  {Srivastava}(2018)}]{lu2018level}%
  \BibitemOpen
  \bibfield  {author} {\bibinfo {author} {\bibfnamefont {Y.}~\bibnamefont
  {Lu}}\ and\ \bibinfo {author} {\bibfnamefont {A.}~\bibnamefont
  {Srivastava}},\ }\bibfield  {title} {\enquote {\bibinfo {title} {Level
  repulsion and band sorting in phononic crystals},}\ }\href@noop {} {\bibfield
   {journal} {\bibinfo  {journal} {Journal of the Mechanics and Physics of
  Solids}\ }\textbf {\bibinfo {volume} {111}},\ \bibinfo {pages} {100--112}
  (\bibinfo {year} {2018})}\BibitemShut {NoStop}%
\bibitem [{\citenamefont {Zhang}\ \emph {et~al.}(2023)\citenamefont {Zhang},
  \citenamefont {Zangeneh-Nejad}, \citenamefont {Chen}, \citenamefont {Lu},\
  and\ \citenamefont {Christensen}}]{zhang2023second}%
  \BibitemOpen
  \bibfield  {author} {\bibinfo {author} {\bibfnamefont {X.}~\bibnamefont
  {Zhang}}, \bibinfo {author} {\bibfnamefont {F.}~\bibnamefont
  {Zangeneh-Nejad}}, \bibinfo {author} {\bibfnamefont {Z.-G.}\ \bibnamefont
  {Chen}}, \bibinfo {author} {\bibfnamefont {M.-H.}\ \bibnamefont {Lu}}, \ and\
  \bibinfo {author} {\bibfnamefont {J.}~\bibnamefont {Christensen}},\
  }\bibfield  {title} {\enquote {\bibinfo {title} {A second wave of topological
  phenomena in photonics and acoustics},}\ }\href@noop {} {\bibfield  {journal}
  {\bibinfo  {journal} {Nature}\ }\textbf {\bibinfo {volume} {618}},\ \bibinfo
  {pages} {687--697} (\bibinfo {year} {2023})}\BibitemShut {NoStop}%
\bibitem [{\citenamefont {Chen}\ \emph {et~al.}(2019)\citenamefont {Chen},
  \citenamefont {Yao}, \citenamefont {Nassar},\ and\ \citenamefont
  {Huang}}]{chen2019mechanical}%
  \BibitemOpen
  \bibfield  {author} {\bibinfo {author} {\bibfnamefont {H.}~\bibnamefont
  {Chen}}, \bibinfo {author} {\bibfnamefont {L.}~\bibnamefont {Yao}}, \bibinfo
  {author} {\bibfnamefont {H.}~\bibnamefont {Nassar}}, \ and\ \bibinfo {author}
  {\bibfnamefont {G.}~\bibnamefont {Huang}},\ }\bibfield  {title} {\enquote
  {\bibinfo {title} {Mechanical quantum hall effect in time-modulated elastic
  materials},}\ }\href@noop {} {\bibfield  {journal} {\bibinfo  {journal}
  {Physical Review Applied}\ }\textbf {\bibinfo {volume} {11}},\ \bibinfo
  {pages} {044029} (\bibinfo {year} {2019})}\BibitemShut {NoStop}%
\bibitem [{\citenamefont {Wang}, \citenamefont {Wang},\ and\ \citenamefont
  {Ma}(2022)}]{wang2022non}%
  \BibitemOpen
  \bibfield  {author} {\bibinfo {author} {\bibfnamefont {W.}~\bibnamefont
  {Wang}}, \bibinfo {author} {\bibfnamefont {X.}~\bibnamefont {Wang}}, \ and\
  \bibinfo {author} {\bibfnamefont {G.}~\bibnamefont {Ma}},\ }\bibfield
  {title} {\enquote {\bibinfo {title} {Non-hermitian morphing of topological
  modes},}\ }\href@noop {} {\bibfield  {journal} {\bibinfo  {journal} {Nature}\
  }\textbf {\bibinfo {volume} {608}},\ \bibinfo {pages} {50--55} (\bibinfo
  {year} {2022})}\BibitemShut {NoStop}%
\bibitem [{\citenamefont {Pernas-Salom{\'o}n}\ and\ \citenamefont
  {Shmuel}(2020)}]{PernasSalomon2019JMPS}%
  \BibitemOpen
  \bibfield  {author} {\bibinfo {author} {\bibfnamefont {R.}~\bibnamefont
  {Pernas-Salom{\'o}n}}\ and\ \bibinfo {author} {\bibfnamefont
  {G.}~\bibnamefont {Shmuel}},\ }\bibfield  {title} {\enquote {\bibinfo {title}
  {Symmetry breaking creates electro-momentum coupling in piezoelectric
  metamaterials},}\ }\href {\doibase
  https://doi.org/10.1016/j.jmps.2019.103770} {\bibfield  {journal} {\bibinfo
  {journal} {Journal of the Mechanics and Physics of Solids}\ }\textbf
  {\bibinfo {volume} {134}},\ \bibinfo {pages} {103770} (\bibinfo {year}
  {2020})}\BibitemShut {NoStop}%
\bibitem [{\citenamefont {Pernas-Salom{\'o}n}\ \emph
  {et~al.}(2021)\citenamefont {Pernas-Salom{\'o}n}, \citenamefont {Haberman},
  \citenamefont {Norris},\ and\ \citenamefont {Shmuel}}]{rps20201wm}%
  \BibitemOpen
  \bibfield  {author} {\bibinfo {author} {\bibfnamefont {R.}~\bibnamefont
  {Pernas-Salom{\'o}n}}, \bibinfo {author} {\bibfnamefont {M.~R.}\ \bibnamefont
  {Haberman}}, \bibinfo {author} {\bibfnamefont {A.~N.}\ \bibnamefont
  {Norris}}, \ and\ \bibinfo {author} {\bibfnamefont {G.}~\bibnamefont
  {Shmuel}},\ }\bibfield  {title} {\enquote {\bibinfo {title} {The
  electromomentum effect in piezoelectric willis scatterers},}\ }\href
  {\doibase https://doi.org/10.1016/j.wavemoti.2021.102797} {\bibfield
  {journal} {\bibinfo  {journal} {Wave Motion}\ ,\ \bibinfo {pages} {102797}}
  (\bibinfo {year} {2021})}\BibitemShut {NoStop}%
\bibitem [{\citenamefont {Muhafra}\ \emph {et~al.}(2022)\citenamefont
  {Muhafra}, \citenamefont {Kosta}, \citenamefont {Torrent}, \citenamefont
  {Pernas-Salom{\'o}n},\ and\ \citenamefont {Shmuel}}]{muhafra2021}%
  \BibitemOpen
  \bibfield  {author} {\bibinfo {author} {\bibfnamefont {A.}~\bibnamefont
  {Muhafra}}, \bibinfo {author} {\bibfnamefont {M.}~\bibnamefont {Kosta}},
  \bibinfo {author} {\bibfnamefont {D.}~\bibnamefont {Torrent}}, \bibinfo
  {author} {\bibfnamefont {R.}~\bibnamefont {Pernas-Salom{\'o}n}}, \ and\
  \bibinfo {author} {\bibfnamefont {G.}~\bibnamefont {Shmuel}},\ }\bibfield
  {title} {\enquote {\bibinfo {title} {Homogenization of piezoelectric planar
  willis materials undergoing antiplane shear},}\ }\href {\doibase
  https://doi.org/10.1016/j.wavemoti.2021.102833} {\bibfield  {journal}
  {\bibinfo  {journal} {Wave Motion}\ }\textbf {\bibinfo {volume} {108}},\
  \bibinfo {pages} {102833} (\bibinfo {year} {2022})}\BibitemShut {NoStop}%
\bibitem [{\citenamefont {Kosta}\ \emph {et~al.}(2022)\citenamefont {Kosta},
  \citenamefont {Muhafra}, \citenamefont {Pernas-Sal{\'o}mon}, \citenamefont
  {Shmuel},\ and\ \citenamefont {Amir}}]{KOSTA2022ijss}%
  \BibitemOpen
  \bibfield  {author} {\bibinfo {author} {\bibfnamefont {M.}~\bibnamefont
  {Kosta}}, \bibinfo {author} {\bibfnamefont {A.}~\bibnamefont {Muhafra}},
  \bibinfo {author} {\bibfnamefont {R.}~\bibnamefont {Pernas-Sal{\'o}mon}},
  \bibinfo {author} {\bibfnamefont {G.}~\bibnamefont {Shmuel}}, \ and\ \bibinfo
  {author} {\bibfnamefont {O.}~\bibnamefont {Amir}},\ }\bibfield  {title}
  {\enquote {\bibinfo {title} {Maximizing the electromomentum coupling in
  piezoelectric laminates},}\ }\href {\doibase
  https://doi.org/10.1016/j.ijsolstr.2022.111909} {\bibfield  {journal}
  {\bibinfo  {journal} {International Journal of Solids and Structures}\
  }\textbf {\bibinfo {volume} {254-255}},\ \bibinfo {pages} {111909} (\bibinfo
  {year} {2022})}\BibitemShut {NoStop}%
\bibitem [{\citenamefont {Muhafra}, \citenamefont {Haberman},\ and\
  \citenamefont {Shmuel}(2023)}]{Muhafra2023PRApplied}%
  \BibitemOpen
  \bibfield  {author} {\bibinfo {author} {\bibfnamefont {K.}~\bibnamefont
  {Muhafra}}, \bibinfo {author} {\bibfnamefont {M.~R.}\ \bibnamefont
  {Haberman}}, \ and\ \bibinfo {author} {\bibfnamefont {G.}~\bibnamefont
  {Shmuel}},\ }\bibfield  {title} {\enquote {\bibinfo {title} {Discrete
  one-dimensional models for the electromomentum coupling},}\ }\href {\doibase
  10.1103/PhysRevApplied.20.014042} {\bibfield  {journal} {\bibinfo  {journal}
  {Phys. Rev. Appl.}\ }\textbf {\bibinfo {volume} {20}},\ \bibinfo {pages}
  {014042} (\bibinfo {year} {2023})}\BibitemShut {NoStop}%
\bibitem [{\citenamefont {Shmuel}\ \emph {et~al.}(2022)\citenamefont {Shmuel},
  \citenamefont {René}, \citenamefont {Alan}, \citenamefont {Majd},
  \citenamefont {Daniel}, \citenamefont {R},\ and\ \citenamefont
  {N}}]{shmuel2022eto}%
  \BibitemOpen
  \bibfield  {author} {\bibinfo {author} {\bibfnamefont {G.}~\bibnamefont
  {Shmuel}}, \bibinfo {author} {\bibfnamefont {P.-S.}\ \bibnamefont {René}},
  \bibinfo {author} {\bibfnamefont {M.}~\bibnamefont {Alan}}, \bibinfo {author}
  {\bibfnamefont {K.}~\bibnamefont {Majd}}, \bibinfo {author} {\bibfnamefont
  {T.}~\bibnamefont {Daniel}}, \bibinfo {author} {\bibfnamefont {H.~M.}\
  \bibnamefont {R}}, \ and\ \bibinfo {author} {\bibfnamefont {N.~A.}\
  \bibnamefont {N}},\ }\bibfield  {title} {\enquote {\bibinfo {title} {Beyond
  willis materials: Trianisotropy and the electromomentum effect},}\
  }\href@noop {} {\bibfield  {journal} {\bibinfo  {journal} {12th International
  Conference on Elastic, Electrical, Transport, and Optical Properties of
  Inhomogeneous Media}\ } (\bibinfo {year} {2022})}\BibitemShut {NoStop}%
\bibitem [{\citenamefont {Danawe}\ and\ \citenamefont
  {Tol}(2023)}]{Danawe2023APL}%
  \BibitemOpen
  \bibfield  {author} {\bibinfo {author} {\bibfnamefont {H.}~\bibnamefont
  {Danawe}}\ and\ \bibinfo {author} {\bibfnamefont {S.}~\bibnamefont {Tol}},\
  }\bibfield  {title} {\enquote {\bibinfo {title} {Electro-momentum coupling
  tailored in piezoelectric metamaterials with resonant shunts},}\ }\href
  {\doibase 10.1063/5.0165267} {\bibfield  {journal} {\bibinfo  {journal} {APL
  Materials}\ }\textbf {\bibinfo {volume} {11}},\ \bibinfo {pages} {091118}
  (\bibinfo {year} {2023})},\ \Eprint
  {http://arxiv.org/abs/https://pubs.aip.org/aip/apm/article-pdf/doi/10.1063/5.0165267/18141427/091118\_1\_5.0165267.pdf}
  {https://pubs.aip.org/aip/apm/article-pdf/doi/10.1063/5.0165267/18141427/091118\_1\_5.0165267.pdf}
  \BibitemShut {NoStop}%
\bibitem [{\citenamefont {Wallen}\ \emph {et~al.}(2022)\citenamefont {Wallen},
  \citenamefont {Casali}, \citenamefont {Goldsberry},\ and\ \citenamefont
  {Haberman}}]{wallen2022polarizability}%
  \BibitemOpen
  \bibfield  {author} {\bibinfo {author} {\bibfnamefont {S.~P.}\ \bibnamefont
  {Wallen}}, \bibinfo {author} {\bibfnamefont {M.~A.}\ \bibnamefont {Casali}},
  \bibinfo {author} {\bibfnamefont {B.~M.}\ \bibnamefont {Goldsberry}}, \ and\
  \bibinfo {author} {\bibfnamefont {M.~R.}\ \bibnamefont {Haberman}},\
  }\bibfield  {title} {\enquote {\bibinfo {title} {Polarizability of
  electromomentum coupled scatterers},}\ }in\ \href {\doibase
  10.1121/2.0001597} {\emph {\bibinfo {booktitle} {Proceedings of Meetings on
  Acoustics}}},\ Vol.~\bibinfo {volume} {46}\ (\bibinfo {organization} {AIP
  Publishing},\ \bibinfo {year} {2022})\BibitemShut {NoStop}%
\bibitem [{\citenamefont {Lee}, \citenamefont {Zhang},\ and\ \citenamefont
  {Gu}(2023)}]{Lee2023JASA}%
  \BibitemOpen
  \bibfield  {author} {\bibinfo {author} {\bibfnamefont {J.-H.}\ \bibnamefont
  {Lee}}, \bibinfo {author} {\bibfnamefont {Z.}~\bibnamefont {Zhang}}, \ and\
  \bibinfo {author} {\bibfnamefont {G.~X.}\ \bibnamefont {Gu}},\ }\bibfield
  {title} {\enquote {\bibinfo {title} {Reaching new levels of wave scattering
  via piezoelectric metamaterials and electro-momentum coupling},}\ }\href
  {\doibase 10.1121/10.0018518} {\bibfield  {journal} {\bibinfo  {journal} {The
  Journal of the Acoustical Society of America}\ }\textbf {\bibinfo {volume}
  {153}},\ \bibinfo {pages} {A163--A163} (\bibinfo {year} {2023})}\BibitemShut
  {NoStop}%
\bibitem [{\citenamefont {Huynh}\ \emph {et~al.}(2023)\citenamefont {Huynh},
  \citenamefont {Zhuang}, \citenamefont {Park}, \citenamefont {Nanthakumar},
  \citenamefont {Jin},\ and\ \citenamefont {Rabczuk}}]{HUYNH2023EML}%
  \BibitemOpen
  \bibfield  {author} {\bibinfo {author} {\bibfnamefont {H.~D.}\ \bibnamefont
  {Huynh}}, \bibinfo {author} {\bibfnamefont {X.}~\bibnamefont {Zhuang}},
  \bibinfo {author} {\bibfnamefont {H.~S.}\ \bibnamefont {Park}}, \bibinfo
  {author} {\bibfnamefont {S.}~\bibnamefont {Nanthakumar}}, \bibinfo {author}
  {\bibfnamefont {Y.}~\bibnamefont {Jin}}, \ and\ \bibinfo {author}
  {\bibfnamefont {T.}~\bibnamefont {Rabczuk}},\ }\bibfield  {title} {\enquote
  {\bibinfo {title} {Maximizing electro-momentum coupling in generalized 2d
  willis metamaterials},}\ }\href {\doibase
  https://doi.org/10.1016/j.eml.2023.101981} {\bibfield  {journal} {\bibinfo
  {journal} {Extreme Mechanics Letters}\ }\textbf {\bibinfo {volume} {61}},\
  \bibinfo {pages} {101981} (\bibinfo {year} {2023})}\BibitemShut {NoStop}%
\bibitem [{\citenamefont {Carcione}(2001)}]{carcione2001wave}%
  \BibitemOpen
  \bibfield  {author} {\bibinfo {author} {\bibfnamefont {J.}~\bibnamefont
  {Carcione}},\ }\href {https://books.google.co.il/books?id=feOC0H1m5_wC}
  {\emph {\bibinfo {title} {Wave Fields in Real Media: Wave Propagation in
  Anisotropic, Anelastic and Porous Media}}},\ ISSN\ (\bibinfo  {publisher}
  {Elsevier Science},\ \bibinfo {year} {2001})\BibitemShut {NoStop}%
\bibitem [{\citenamefont {Srivastava}\ and\ \citenamefont
  {Willis}(2017)}]{Srivastava2017PRA}%
  \BibitemOpen
  \bibfield  {author} {\bibinfo {author} {\bibfnamefont {A.}~\bibnamefont
  {Srivastava}}\ and\ \bibinfo {author} {\bibfnamefont {J.~R.}\ \bibnamefont
  {Willis}},\ }\bibfield  {title} {\enquote {\bibinfo {title} {{Evanescent wave
  boundary layers in metamaterials and sidestepping them through a variational
  approach}},}\ }\href {\doibase 10.1098/rspa.2016.0765} {\bibfield  {journal}
  {\bibinfo  {journal} {Proc. R. Soc. London A Math. Phys. Eng. Sci.}\ }\textbf
  {\bibinfo {volume} {473}} (\bibinfo {year} {2017}),\
  10.1098/rspa.2016.0765}\BibitemShut {NoStop}%
\bibitem [{\citenamefont {Mokhtari}, \citenamefont {Lu},\ and\ \citenamefont
  {Srivastava}(2019)}]{mokhtari2019properties}%
  \BibitemOpen
  \bibfield  {author} {\bibinfo {author} {\bibfnamefont {A.~A.}\ \bibnamefont
  {Mokhtari}}, \bibinfo {author} {\bibfnamefont {Y.}~\bibnamefont {Lu}}, \ and\
  \bibinfo {author} {\bibfnamefont {A.}~\bibnamefont {Srivastava}},\ }\bibfield
   {title} {\enquote {\bibinfo {title} {On the properties of phononic
  eigenvalue problems},}\ }\href@noop {} {\bibfield  {journal} {\bibinfo
  {journal} {Journal of the Mechanics and Physics of Solids}\ }\textbf
  {\bibinfo {volume} {131}},\ \bibinfo {pages} {167--179} (\bibinfo {year}
  {2019})}\BibitemShut {NoStop}%
\bibitem [{\citenamefont {Khodavirdi}, \citenamefont {Mokhtari},\ and\
  \citenamefont {Srivastava}(2022)}]{khodavirdi2022scattering}%
  \BibitemOpen
  \bibfield  {author} {\bibinfo {author} {\bibfnamefont {H.}~\bibnamefont
  {Khodavirdi}}, \bibinfo {author} {\bibfnamefont {A.~A.}\ \bibnamefont
  {Mokhtari}}, \ and\ \bibinfo {author} {\bibfnamefont {A.}~\bibnamefont
  {Srivastava}},\ }\bibfield  {title} {\enquote {\bibinfo {title} {Scattering
  of mechanical waves from the perspective of open systems},}\ }\href@noop {}
  {\bibfield  {journal} {\bibinfo  {journal} {Mechanics of Materials}\ }\textbf
  {\bibinfo {volume} {172}},\ \bibinfo {pages} {104399} (\bibinfo {year}
  {2022})}\BibitemShut {NoStop}%
\bibitem [{\citenamefont {Khodavirdi}, \citenamefont {Ong},\ and\ \citenamefont
  {Srivastava}(2023)}]{khodavirdi2023atomistic}%
  \BibitemOpen
  \bibfield  {author} {\bibinfo {author} {\bibfnamefont {H.}~\bibnamefont
  {Khodavirdi}}, \bibinfo {author} {\bibfnamefont {Z.-Y.}\ \bibnamefont {Ong}},
  \ and\ \bibinfo {author} {\bibfnamefont {A.}~\bibnamefont {Srivastava}},\
  }\bibfield  {title} {\enquote {\bibinfo {title} {The atomistic green's
  function method for acoustic and elastic wave-scattering problems},}\
  }\href@noop {} {\bibfield  {journal} {\bibinfo  {journal} {arXiv preprint
  arXiv:2301.12259}\ } (\bibinfo {year} {2023})}\BibitemShut {NoStop}%
\bibitem [{\citenamefont {Zhou}\ \emph {et~al.}(2024)\citenamefont {Zhou},
  \citenamefont {Jiang}, \citenamefont {Zhu}, \citenamefont {Li}, \citenamefont
  {Li}, \citenamefont {Wang}, \citenamefont {Qiu},\ and\ \citenamefont
  {Wang}}]{zhou2024underwater}%
  \BibitemOpen
  \bibfield  {author} {\bibinfo {author} {\bibfnamefont {H.-T.}\ \bibnamefont
  {Zhou}}, \bibinfo {author} {\bibfnamefont {M.}~\bibnamefont {Jiang}},
  \bibinfo {author} {\bibfnamefont {J.-H.}\ \bibnamefont {Zhu}}, \bibinfo
  {author} {\bibfnamefont {Y.}~\bibnamefont {Li}}, \bibinfo {author}
  {\bibfnamefont {Q.}~\bibnamefont {Li}}, \bibinfo {author} {\bibfnamefont
  {Y.-F.}\ \bibnamefont {Wang}}, \bibinfo {author} {\bibfnamefont {C.-W.}\
  \bibnamefont {Qiu}}, \ and\ \bibinfo {author} {\bibfnamefont {Y.-S.}\
  \bibnamefont {Wang}},\ }\bibfield  {title} {\enquote {\bibinfo {title}
  {Underwater scattering exceptional point by metasurface with fluid-solid
  interaction},}\ }\href@noop {} {\bibfield  {journal} {\bibinfo  {journal}
  {Advanced Functional Materials}\ ,\ \bibinfo {pages} {2404282}} (\bibinfo
  {year} {2024})}\BibitemShut {NoStop}%
\end{thebibliography}%

\end{document}